\documentclass[conference,letterpaper]{IEEEtran}
\usepackage[utf8]{inputenc} 
\usepackage[T1]{fontenc}
\usepackage{url}
\usepackage{ifthen}
\usepackage{cite}
\usepackage[cmex10]{amsmath} % Use the [cmex10] option to ensure complicance
\usepackage{balance}
\usepackage{color,xcolor}
\usepackage{subfigure}  %插入多图时用子图显示的宏包
\usepackage{verbatim} % 添加评论用
\newtheorem{Theo}{\bf Theorem}
\newtheorem{Lem}{\bf Lemma}
\newtheorem{Cor}{\bf Corollary}
\newtheorem{remark}{Remark}
\allowdisplaybreaks % 允许换行
\usepackage{multirow}%提供跨行列命令\multicolumn{}{}{}
\usepackage{epsfig}
\usepackage{amssymb} % 某些字体
\usepackage{makecell}

\begin{document}
	\title{Demand Private Coded Caching: the Two-File Case} 
	\author{
		\IEEEauthorblockN{
			Qinyi Lu\IEEEauthorrefmark{*},
			Nan Liu\IEEEauthorrefmark{+} and
			Wei Kang\IEEEauthorrefmark{*}}
		\IEEEauthorblockA{\IEEEauthorrefmark{*}School of Information Science and Engineering, Southeast University, Nanjing, China 210096}
		\IEEEauthorblockA{\IEEEauthorrefmark{+}National Mobile Communications Research Laboratory, Southeast University, Nanjing, China 210096}
		\IEEEauthorblockA{qylu@seu.edu.cn,nanliu@seu.edu.cn,wkang@seu.edu.cn}}
	
	\DeclareRobustCommand*{\IEEEauthorrefmark}[1]{%作者标号是数字
		\raisebox{0pt}[0pt][0pt]{\textsuperscript{\footnotesize\ensuremath{#1}}}}

	\maketitle
	%%%%%%
	%% Abstract: 
	%% If your paper is eligible for the student paper award, please add
	%% the comment "THIS PAPER IS ELIGIBLE FOR THE STUDENT PAPER
	%% AWARD." as a first line in the abstract. 
	%% For the final version of the accepted paper, please do not forget
	%% to remove this comment!
	%%
	\begin{abstract}
		% THIS PAPER IS ELIGIBLE FOR THE STUDENT PAPER AWARD.
		We investigate the demand private coded caching problem, which is an $(N,K)$ coded caching problem with $N$ files, $K$ users, each equipped with a cache of size $M$, and an additional privacy constraint on user demands. 
		We first present a new virtual-user-based achievable scheme for arbitrary number of users and files. Then, for the case of 2 files and arbitrary number of users, we derive some new converse bounds. As a result, we obtain the exact memory-rate tradeoff of the demand private coded caching problem for 2 files and 3 users. As for the case of 2 files and arbitrary number of users, the exact memory-rate tradeoff is characterized  for $M\in [0,\frac{2}{K}] \cup [\frac{2(K-1)}{K+1},2]$. 
	\end{abstract}
	
	\section{Introduction}
	% \vspace{-0.2cm}
	% coded caching 背景(删去)
	Maddah Ali and Niesen proposed  the coded caching problem from the perspective of information theory and studied its fundamental  limits in \cite{MaddahAli2014}. 
	% \begin{comment}
	%% 模型介绍
	% {\color{red}MyQ: In order to compress the layout, I would like to write this section shorter and just need to explain clearly what N, K and M are respectively}
	There are $K$ users and one server, who has access to $N$ files of equal size in the system. Each user has a cache which can store $M$ files and is connected to the server through an error-free shared link. 
	%{\color{red}if you describe the system model, you need to talk about the two phases and the communication demand.}
	% 添加the two phases and the communication demand
	In the placement phase, the cache of each user is filled by a function of $N$ files.
	In the delivery phase, each user requests one file from the server, and the server broadcasts a signal to all $K$ users based on the received requests. 
	The goal is to minimize the rate of the broadcast signal, given a cache size $M$, while ensuring that each user can correctly decode the required file.
	% \end{comment}
	%%
	The caching and delivery scheme  proposed  in  \cite{MaddahAli2014} is proved to be order optimal and 
	the exact memory-rate tradeoff for the coded caching problem is still open for general $N$ and $K$. 
	Several works had made efforts to characterize the exact memory-rate tradeoff by proposing improved achievable schemes \cite{Wan2016, Chen2016, Mohammadi2016, Mohammadi2017, Qian2018, Tian2018,  Gomez2018, Kumar2019, Shao2019ITW, Shao2019Tcom, Wan2020 } or proving tighter bounds \cite{Sengupta2015, Ajaykrishnan2015, Tian2016, Ghasemi2017, Wang2018, Qian2019,Kumar2023}.
	In particular, the scheme proposed by Yu \emph{et al.}, refered to as \emph{the YMA scheme}, % 加了斜体
	is proved to be optimal under the condition of uncoded placement \cite{Qian2018}. 
	
	%% demand privacy 研究现状
	For schemes designed for the traditional coded caching problem, users participating may know the indices of the files requested by other users.
	This is undesirable for participating users as it violates their privacy.
	Several works have investigated the coded caching problem with the constraint of demand privacy.
	More specifically,  Gurjarpadhye  \emph{et al.}  proposed three schemes and the best performance of these three schemes together is shown to be order optimal for the demand private coded caching problem\cite{Chinmay2022}. 
	 In particular, the scheme stated in \cite[Theorem 2]{Chinmay2022} is a virtual user scheme based on an $(N,NK,M,R)$  non-private scheme with restricted demand subset stated in \cite[Lemma 1]{Chinmay2022}. 
	We call the scheme in \cite[Lemma 1]{Chinmay2022} \emph{the GRK non-private scheme}, which is a coded caching scheme designed without considering demand privacy, where there are $N$ files, $NK$ users  and the demands are restricted to a set of special  demands of size $N^K$. 
	The idea of \emph{virtual user} ensures user's demand privacy by creating $NK-K$ virtual users and having each file requested by exactly $K$ users out of a total of $NK$ users.
	In addition, the exact memory-rate tradeoff with $N \ge K = 2$ for the problem is found\cite{Chinmay2022}. 
	In \cite{Aravind2020, Aravind2022}, Aravind \emph{et al.} mainly focused on  reducing subpacketization  and as a result, they proposed new schemes using placement delivery arrays (PDAs). 
	In \cite{Wan2021}, Wan and Caire considered the demand private coded caching problem with multiple requests and  characterized the exact optimality when $M \ge \min \left\{ \frac{2K-1}{2K}, \frac{2^{K-1}}{2^{K-1}+1} \right\} N $.
	In \cite{Yan2021}, Yan and  Tuninetti considered demand privacy against colluding users in both scenarios: Single File Retrieval (SFR) and Linear Function Retrieval (LFR), and provided a new converse bound for SFR scenario.
	The exact memory-rate tradeoff when $ M \le \frac{1}{K(N-1)+1}$ for the SFR scenario is characterized in \cite{Namboodiri2021}.
	In \cite{Gholami2023}, Gholami \emph{et al.} constructed a new private coded caching scheme based on any
	%别用through 
	Private Information Retrieval (PIR)  scheme and investigated the coded caching problem with simultaneously private demands and caches. 
	%% 感觉这些模型可以不引用
	Several demand private problems under different coded caching models such as Multi-access\cite{Wan2021Multi, Namboodiri2022, Chinnapadamala2022}, Device-to-Device \cite{Wan2020D2D, Wan2020D2DConverse, Wan2022D2D}, with a wiretapper\cite{Yan2021Security, Yan2022Secure} and  with offline users \cite{Yan2022Secure,Ma2023, ma2024coded} have also been investigated. 
	%, ma2024coded
	Just like the coded caching problem without privacy constraint, the exact memory-rate tradeoff for general demand private coded caching problem is still open.
	
	In this paper, we focus on the demand private coded caching problem studied in \cite{Chinmay2022}.
	We first present a new achievable scheme for arbitrary number of users and files.
	The proposed scheme is based on the idea of virtual user, and its design is related to the YMA scheme proposed in \cite{Qian2018}.
	Numerical results show that the proposed scheme outperforms existing achievable schemes when the cache size $M$ is small and $N \leq K$. 
	Next, for the case of two files and arbitrary number of users, we derive some new converse results, which are obtained using techniques such as induction and recursion.  
	%by combining certain entropy terms of different cache contents and transmitted signals through the submodularity of the entropy function, applying the decoding constraint, and exploiting the equality of certain entropy terms caused by the privacy constraint.  The methods of the proof include induction and recursive proof. 
	Finally, for the case of 2 files and 3 users, we show that the new proposed achievable scheme, together with the achievability result in \cite[Theorem 2]{Chinmay2022}, and the proposed converse meet,
	thus characterizing the exact memory-rate tradeoff.
	As for the case of 2 files and arbitrary number of users, the exact memory-rate tradeoff for the case of $M\in [0,\frac{2}{K}] \cup [\frac{2(K-1)}{K+1},2]$ is obtained.
	
	% The rest of the paper is organized as follows.
	%%%%%%%%%%%%%%%%%% 符号约定
	\emph{Notations:}
	% We use $|\mathcal{A}|$ to denote the cardinality of a set $\mathcal{A}$,  
	We use $[a:b]$ to denote the set $\{a,a+1,\dots,b\}$,  % t$.
	%% {\color{blue} $[a]=[1:a]$ to denote the set  $\{1,\dots,a\}$},
	%% and $[a, b]$ to denote the closed interval between two real numbers $a$ and $b$.
	and $X_{\mathcal{I}} :=  \left( X_i\right)_{i\in \mathcal{I} }$ to denote a set of random variables.
	% i.e., $ a \oplus_N b = (a+b)\mod N $.
	% Bold symbols denote vectors;  
	$\boldsymbol{a}(j)$ denotes the $j$-th bit of vector $\boldsymbol{a}$ and $j$ starts from $0$; % 向量的第j位(注意本文从第0位开始算)
	$\mathbf{e}_k \in \mathbb{F}^{NK-K+1}_2$ denotes the unit vector where the 
	$k$-th element is 1 and the other elements are 0;
	% 建议：1K 乘上系数 
	$\boldsymbol{1}_K:= (1,\dots,1)$  denotes the $K$ length vector with all elements equal to $1$, and
	$(a\boldsymbol{1}_{K-k},b\boldsymbol{1}_k)$ denotes the $K$ length vector composed of $K-k$ elements of $a$ followed by $k$ elements of $b$.
	% $\oplus$ denotes bit-wise XOR; 
	% and $\oplus_N$  denotes the integer addition modulo $N$.
	Let $\tilde{a} = (a+1) \mod 2$  for $a = 0,1$, and we assume that $\sum_{i=a}^b \cdot  = 0$ when $a > b$. 
	
	\section{System Model}
	We study an $(N,K)$ demand private coded caching problem as follows. The system has a
	server connected to $K$  cache-aided users through an error-free shared link.
	The server has access to a library of $N$ independent files of $F$ bits each. Let  $(W_0,W_1,\dots,W_{N-1})$ represent these files and these files are uniformly distributed on $\{0,1\}^F$. Each user $k\in[0:K-1]$ has a  cache whose size is limited to  $MF$ bits. 
	The coded caching system operates in two phases as follows.
	\subsubsection{Placement Phase} % 这里加入key
	From the probability space $ \mathcal{P}$, the server generates a random variable $P$, 
	% the  random variable $P$ is not known to any user.
	which is not known to any user.
	Then the server places at most $MF$ bits in user $k$'s cache through a cache function $\phi_k$. More specifically, for $\forall k \in [0:K-1]$, the cache function of the $k$-th user is  known to  all $K$ users in the system and is a map  as follows 
	\begin{align*}
	\phi_k: \mathcal{P} \times [0:2^F-1]^N \longrightarrow [0:2^{MF}-1] .
	\end{align*} 
	Let $Z_k$ denote the cache content for user $k$, so $Z_k$ is given by 
	\begin{align}   \label{sysZ}
	Z_k = \phi_k \left( W_{[0:N-1]},P \right), \quad \forall k\in[0:K-1].
	\end{align}
	
	\subsubsection{Delivery Phase}
	In the delivery phase, each user demands one of the $N$ files. 
	Let $D_k$ represent the demand of user $k$, and the demand for all users in the system can be denoted by the vector $\boldsymbol{D} = (D_0,D_1, \dots, D_{K-1})$. 
	$D_{[0:K-1]}$ are all i.i.d. random variables and uniformly
	distributed on $\{0,1,\dots ,N-1\} $.
	The files $W_{[0:N-1]}$, the randomness  $P$ and the demands $\boldsymbol{D}$ are independent, i.e., 
	\begin{align}  \label{sysindependent}
	H\left( \boldsymbol{D}, P, W_{[0:N-1]} \right) = \textstyle \sum_{k=0}^{K-1} H(D_k) + H(P) + NF. 
	\end{align} 
	Note that the demands $\boldsymbol{D}$ are also independent of the cache contents $Z_{[0:K-1]}$. As $Z_k$ is  a deterministic function of $W_{[0:N-1]}$ and $P$ (see \eqref{sysZ}), the independence between $\boldsymbol{D}$ and $Z_{[0:K-1]}$ is implied by \eqref{sysindependent}.
	Each user sends its own demand to the server through a  private link. Upon receiving the demands of all users, the server broadcasts a signal $X_{\boldsymbol{D}}$  of size $RF$ bits and the signal $X_{\boldsymbol{D}}$ is created through an encoding function $\psi$. 
	The encoding function $\psi$ is a mapping as follows 
	\begin{align*}
	\psi : \mathcal{P} \times [0:N-1]^K \times [0:2^F-1]^N \longrightarrow [0:2^{RF}-1],
	\end{align*} 
	% Let $X_{\boldsymbol{D}}$  denote the delivery signal under $\boldsymbol{D}$, so $X_{\boldsymbol{D}}$ is given by
	where $R$ is referred to as the rate of the shared link.
	Hence, the delivery signal under demand $\boldsymbol{D}$, i.e., $X_{\boldsymbol{D}}$, is given by
	\begin{align*} % \label{sysX}
	X_{\boldsymbol{D}} = \psi \left(W_{[0:N-1]},\boldsymbol{D},P  \right).  
	\end{align*} 
	The $(N,K,M,R)$-private coded caching scheme consists of $\phi_{[0:K-1]}$ and $ \psi$, and it must  satisfy the following correctness and privacy constraints.
	
	\emph{Correctness:}
	Each user $k\in[0:K-1]$ must decode its  demand file $W_{D_k}$ by using $(X_{\boldsymbol{D}},Z_k)$ with no error, i.e., 
	\begin{align}  \label{decoding}
	H\left( W_{D_k} | Z_k, X_{\boldsymbol{D}},D_k\right) = 0, \quad \forall k\in [0:K-1].
	\end{align} 
	
	\emph{Privacy:}
	The demand privacy requires that any user can not infer any information about the demands $\boldsymbol{D}$ beyond its own demand $D_k$ from all the information it possesses, i.e., 
	\begin{align}  \label{privacy2}
	I \left(D_{[0:K-1]\setminus \{k\}};X_{\boldsymbol{D}},Z_{k} ,D_{k} \right) = 0, 
	\quad \forall k \in [0:K-1].
	\end{align}
	
	The pair $(M,R)$ is said to be achievable for the $(N,K)$ coded caching problem with demand privacy if there exists an $(N,K,M,R)$-private coded caching scheme for large enough $F$.
	The optimal memory-rate tradeoff for the $(N,K)$ coded caching problem with demand privacy is defined as 
	\begin{align*}  %\label{stsMR} 
	R_{N,K}^{*p}(M) = \inf \{R: \text{$(M,R)$  is achievable} \}. 
	\end{align*}
	where the infimum is taken over all achievable $(N,K,M,R)$-private coded caching schemes. 
	
	\section{Main Result and Discussions}
	The following are the main results  of this paper.  First, we provide a new achievability result for arbitrary number of files and users in the following theorem.
	%Theorem \ref{ach2} states {\color{red}the performance of the proposed achievable} scheme. % and the proof of Theorem \ref{ach2} is later given in Section \ref{secach}.
	\begin{Theo}   \label{ach2}% general的flip方案
		For an $(N,K)$ demand private coded caching problem, the lower convex envelope of the following memory-rate pairs
		\begin{align}  
		\left(M,R \right) = &   \left(  \frac{\tbinom{NK-K+1}{r+1}-\tbinom{NK-K-N+1}{r+1}}{\tbinom{NK-K+1}{r}}
		,\frac{Nr}{NK-K+1} \right),  \nonumber 
		\end{align}
		where $r \in [0:NK-K+1] $ is achievable.
	\end{Theo}
	\begin{IEEEproof}  
		We propose a new achievable scheme based on the idea of virtual user, where $NK$ possible cache contents are designed according to the delivery signals under the corresponding demands in the YMA scheme. The delivery signals for $NK-K+1$ out of $N^K$ demands are obtained based on the cache contents in the YMA scheme, and then the delivery signals for the remaining demands are constructed.
		The detailed proof can be found in Section \ref{secach}. 
	\end{IEEEproof} 
	 
	\begin{remark}
		Compared with the scheme proposed in \cite[Theorem 2]{Chinmay2022}, the proposed scheme can be regarded as a virtual user achievable scheme based on a new non-private scheme with a restricted demand subset. Let $M^{\text{GRK}}$ and $R^{\text{GRK}}$ denote the cache size and rate of \emph{the GRK non-private scheme}, the $(M,R)$ pair of the newly proposed non-private scheme satisfies: $(M,R) = (R^{\text{GRK}}, M^{\text{GRK}})$,   i.e., the cache size and delivery signal rate are swapped. 
	\end{remark} 
	\begin{remark}% {\rm (Numerical analysis) }
		Numerical results show that our proposed scheme in Theorem \ref{ach2} is strictly better than existing achievable schemes for small cache and $N \le K$, e.g., $(N,K)=(5,10)$, $M \in [0.03,1.67]$.
		% From the numerical results shown  in Fig. \ref{fig1a}, we observe that our proposed scheme in Theorem \ref{ach2} outperforms known achievable  schemes when the cache size is small {\color{blue}for $(N,K)=(5,10)$.}
		% Furthermore,  when considering the stronger privacy constraint,
		% since schemes B and C in \cite{Chinmay2022} does not satisfy the stronger privacy constraint, our scheme outperforms known achievable  schemes {\color{blue}for $(N,K)=(5,10)$ and $(N,K)=(10,5)$} as shown in Fig. \ref{fig1a} and Fig. \ref{fig1b}.  %
	\end{remark}
	
	% N = 2 的一个converse 
	Next, the following theorem states a converse result for the case of 2 files and arbitrary number of users. % and the proof of this converse will be given in Section \ref{seccon}.
	\begin{Theo} \label{con2}  % general的flip方案
		For the $(2,K)$ demand private coded caching problem, any $(M, R)$ pair must satisfy that for any $k\in[2:K]$,
		\begin{subequations}
			\begin{align}
			(k+1)(k+2) M + 2k(k+1)R &\ge 2k(k+3), \text{ and } \label{con21} \\
			2k(k+1)M + (k+1)(k+2) R &\ge 2k(k+3) \label{con22}.
			\end{align}
		\end{subequations}
	\end{Theo}
	
	\begin{IEEEproof}
		The proof is derived by suitably combining entropy terms containing different cache contents and delivery signals through the submodularity of the entropy function, applying the correctness constraint 
		%in 为了少一行
		\eqref{decoding}, and exploiting the equality of certain entropy terms caused by the privacy constraint 
		%in 
		\eqref{privacy2}. The proof involves the methods of induction and recursion.
		% The content of these entropy terms and their combination order are carefully designed to determine the maximum number of files.
		The detailed proof can be found in Section \ref{seccon}.
	\end{IEEEproof} 
	
	By comparing the achievability results of Theorem 1 and the converse result of Theorem 2, Corollaries \ref{cor1} and \ref{cor2} state some exact memory-rate tradeoff results for  the $(2,K)$ demand private coded caching problem. 
	
	\begin{Cor} \label{cor1}  % N = 2,K>=4 small buffer 和 large  buffer的optimal
		For the $(2,K)$ demand private coded caching problem, 
		%there exists scheme  which is optimal when  $M\in [0,\frac{2}{K}] \cup [\frac{2(K-1)}{K+1},2] $, i.e., 
		when $M\in[0,\frac{2}{K}]$,
		\vspace{-0.1cm}
		\begin{align*}
		R^{*p}_{N,K}(M) = \max \left\{2-2M, \frac{2K(K+3)}{(K+1)(K+2)}-\frac{2K}{K+2}M \right\},
		\end{align*}
		and when $M\in[\frac{2(K-1)}{K+1},2]$,
		\vspace{-0.1cm}
		\begin{align*}
		R^{*p}_{N,K}(M) = \max \left\{1- \frac{1}{2}M,\frac{K+3}{K+1}-\frac{K+2}{2K}M  \right\} .
		\end{align*}
	\end{Cor}
	% 推论1的简单推导  
	\vspace{-0.2cm}
	\begin{Cor} \label{cor2}  % N = 2,K=3的optimal
		For the $(2,3)$ demand private coded caching problem, the exact memory-rate tradeoff is
		\begin{align*} % \label{eqcor1}
		& R_{N,K}^{*p}(M)   \\
		& \quad = \max \Big \{ 2-2M, \frac{9-6M}{5}, \frac{5-3M}{3},  \frac{9-5M}{6} ,  \frac{2-M}{2} \Big \}.
		% 2M + R \ge 2,6M + 5R \ge 9,3M+3R\ge 5, \nonumber \\
		%  5M + 6R \ge 9, M+2R \ge 2.
		\end{align*}
	\end{Cor}
	\begin{IEEEproof}
		Corollary \ref{cor1} and \ref{cor2} can be easily proved by combining the results in Theorems \ref{ach2} and \ref{con2}, \cite[Theorem 2]{Chinmay2022} and the cut-set bound for non-private coded caching problem in \cite{MaddahAli2014}.  The details is in Appendix \ref{pfCoro1} and \ref{pfCoro2}, respectively.
	\end{IEEEproof}
	\begin{remark}
			Corollary \ref{cor1} contains the optimal result characterized in \cite[Theorem 6]{Wan2021} for $N=2$, as $\frac{2(K-1)}{K+1} \le 2 \min   \left\{ \frac{2K-1}{2K}, \frac{2^{K-1}}{2^{K-1}+1} \right\}  $.
			In addition, by setting $K=2$ in Corollary \ref{cor1}, we obtain the exact memory-rate tradeoff of the $(2,2)$ demand private coded caching problem, which was characterized in \cite[Theorem 6]{Chinmay2022}.
	\end{remark}

	\section{Proof of Theorem \ref{ach2}} \label{secach}
	%
	%$(N,NK,M,R)$ $\mathcal{D_{RS}}$-non-private scheme
	%
	%$(N,K,M,R)$-private scheme
	%
	\subsection{Motivating Example: $N=2,K=3,r=2$}\label{pfachex}		
   	We describe the proposed scheme for the $(2,3)$ demand private coded caching problem achieving $(M,R) = (\frac{2}{3},1)$. 
	
	\emph{Placement Phase}. 
		Split file $n=0,1$ into $6$ equal size subfiles, i.e., $W_n = \left(W_{n, \mathcal{R}}  \right)_{\mathcal{R}\subset [0:3], |\mathcal{R} |=2}$.  
		The server randomly chooses either $Z_{k,0}$ or $Z_{k,1}$ with equal probability to place in user $k$'s  cache.  
		The   choice is denoted by a key $S_k$,
		which is known only to the server and user $k$,  and  follows a uniform distribution on $\mathbb{F}_2$.
		For each $k = 0,1,2$ and $S_k = 0,1$, we design a  $1 \times 4$ demand vector denoted by $\boldsymbol{u}^{(k,S_k)} = (u_0^{(k,S_k)},u_1^{(k,S_k)},u_2^{(k,S_k)},u_3^{(k,S_k)})$. Table \ref{table2} summarizes the design results.
		\begin{table}[h!] 
			\vspace{-0.5cm}
			\setlength{\abovecaptionskip}{0cm} % 调整间距
			\setlength{\belowcaptionskip}{-1cm}
			% \small
			\begin{center}
				\caption{The design of $\boldsymbol{u}^{(k,S_k)}$ in  $ N=2,K=3 $ case}  \label{table2}
				\begin{tabular}{|c|c|c|c|}
					\hline
					$ \boldsymbol{u}^{(k,S_k)} $   &   $ k = 0$ &   $ k = 1$ &   $ k = 2$\\
					\hline
					$ S_k = 0 $   &   $ (0,1,0,0)$ &  $(0,1,0,1)$  &   $(0,1,1,1)$\\
					\hline
					$ S_k = 1 $   &  $ (1,0,1,1)$ &   $(1,0,1,0)$ &   $(1,0,0,0)$ \\
					\hline
				\end{tabular}
			\end{center}
			\vspace{-0.3cm}
		\end{table}
		Thus, the cache content of  user $k$, i.e., $Z_{k}$, is given by 
		$
		Z_k  =  (Z_{k,S_k},S_k),   
		$
		where % 去掉main payload的表述
		% the possible main payload {\color{red} is payload the right term here?}, i.e., 
		$ Z_{k,S_k}$ can be constructed through the delivery signal in the YMA scheme under demand $\boldsymbol{u}^{(k,S_k)}$ when $N=2$ and $K=4$, i.e., 
		\begin{align*}
		& Z_{k,S_k} =  X^{\text{YMA}}_{\boldsymbol{u}^{(k,S_k)}} 
		=   \big( Y_{\mathcal{R}^+}^{(k,S_k)} \big)_{\mathcal{R}^+ \subset [0:3], |\mathcal{R}^+|=3}, \\
		& \text{where} \quad Y_{\mathcal{R}^+}^{(k,S_k)}  =  \bigoplus_{i\in \mathcal{R}^+} W_{u_i^{(k,S_k)}, \mathcal{R}^+ \setminus \{i\}}.
		\end{align*} 
		
		Taking the example of $k=1$ and $S_k=0$, we have $ Z_{1,0} = (Y_{ \{0,1,2\}}^{(1,0)},Y_{ \{0,1,3\}}^{(1,0)},	Y_{ \{0,2,3\}}^{(1,0)}, Y_{ \{1,2,3\}}^{(1,0)})$ where
		\begin{align*}
			Y_{ \{0,1,2\}}^{(1,0)} & =  W_{0,\{1,2\}} \oplus W_{1,\{0,2\}} \oplus  W_{0,\{0,1\}},  \\ 
			Y_{ \{0,1,3\}}^{(1,0)} & =  W_{0,\{1,3\}} \oplus W_{1,\{0,3\}} \oplus  W_{1,\{0,1\}},  \\ 
			Y_{ \{0,2,3\}}^{(1,0)} & =  W_{0,\{2,3\}} \oplus W_{0,\{0,3\}} \oplus  W_{1,\{0,2\}},  \\ 
			Y_{ \{1,2,3\}}^{(1,0)} & =  W_{1,\{2,3\}} \oplus W_{0,\{1,3\}} \oplus  W_{1,\{1,2\}}.  
		\end{align*}
	
	Given that for a sufficiently large file length $F$, the size of $S_k$ can be neglected, and each $Z_{k,S_k}$ contains $4$ segments $Y_{\mathcal{R}^+}^{(k,S_k)}$ of length $\frac{F}{6}$, we can determine that $M = \frac{4F}{6F} = \frac{2}{3}$. 
			
	\emph{Delivery Phase}.  The server constructs an auxiliary  vector $\boldsymbol{d} = (d_0, d_1, d_{2})$ based on the demand  $\boldsymbol{D} = (D_0,D_1,D_2)$ and the keys $\boldsymbol{S} = (S_0, S_1,S_2 )$  as follows
	\begin{align*}
		     (d_0, d_1, d_{2}) = (D_0 \oplus S_0, D_1 \oplus S_1, D_2 \oplus S_2 ).
		     % 在这里general的情况是$\ominus$
	\end{align*}
	Then the server generates the delivery signal $X_{\boldsymbol{D}}$ based on $\boldsymbol{d}$. More specifically, this process involves two steps: first, creating a set $\mathcal{V_{\boldsymbol{d}}}$ and randomly selecting an element from $\mathcal{V_{\boldsymbol{d}}}$, denoted as $t_{\boldsymbol{d}}$; then, based on $\mathcal{V_{\boldsymbol{d}}}$ and $t_{\boldsymbol{d}}$, generating $X_{\boldsymbol{D}}$.
	% $X_{\boldsymbol{D}}  = (X^n_{\boldsymbol{d},\{t\}})_{t \in [0:3] \setminus \{t_d\}, n\in \{0,1\}}$.
	Table \ref{table1} lists the delivery signals under different $\boldsymbol{d}$. Each $X_{\boldsymbol{d},\{t\}}^{n}$ in the table represents the delivery signal containing $X_{\boldsymbol{d},\{t\}}^{0}$ and $X_{\boldsymbol{d},\{t\}}^{1}$, with "/" indicating the absence of that segment in the delivery signal.
	Additionally, the auxiliary vector $\boldsymbol{d}$ is also included in $X_{\boldsymbol{D}}$.
	For a sufficiently large file length $F$, the size of $\boldsymbol{d}$ can be neglected, and as shown in Table \ref{table1}, each $X_{\boldsymbol{D}}$ contains $6$ segments $X_{\boldsymbol{d},\{t\}}^{n}$ of length $\frac{F}{6}$, thus, $M = \frac{6F}{6F} = 1$.
\vspace{0cm}	
\begin{table*}[h!] 
	\setlength{\abovecaptionskip}{0cm} % 调整间距
	\setlength{\belowcaptionskip}{-1cm}
	% \small
	\begin{center}
		\caption{Transmissions of  $( N=2,K=3,M= \frac{2}{3},R= 1)$-private scheme}  \label{table1}
		\begin{tabular}{|c|c|c|c|c|c|c|}
			% 表头
			\hline
			$\boldsymbol{d}$   & $\mathcal{V}_{\boldsymbol{d}}$  & $t_{\boldsymbol{d}} $ & $ X^n_{\boldsymbol{d}, \{0\}}$ &  $ X^n_{\boldsymbol{d}, \{1\}}$ &   $ X^n_{\boldsymbol{d}, \{2\}}$ &  $ X^n_{\boldsymbol{d}, \{3\}}$ \\
			\hline
			% YMA-0
			$(0,0,0) $   & $\{0\}$ & $ 0 $ & $/$
			& $W_{n,\{0,1\}}$ & $ W_{n,\{0,2\}}$ & 
			$W_{n,\{0,3\}}$   \\
			\hline % YMA-1
			$(1,1,1)$ &  $\{1\}$  & $ 1 $ & $W_{n,\{0,1\}}$ & $/$  & $W_{n,\{1,2\}}$  & $W_{n,\{1,3\}}$ \\
			\hline % YMA-2
			$(0,0,1)$ &  $\{2\}$  & $ 2 $ & $W_{n,\{0,2\}}$  &  $W_{n,\{1,2\}}$  &  $/$  &  $W_{n,\{2,3\}}$\\
			\hline % V=012 td = 0 —— 跨行
			$(1,1,0)$ &  $\{0,1,2\}$  & $ 0 $ &  $/$   &  $W_{n,\{0,1\}} \oplus W_{n,\{1,2\}}$  & $W_{n,\{0,2\}}\oplus W_{n,\{1,2\}}$   &  $ W_{n,\{0,3\}} \oplus W_{n,\{1,3\}}  \oplus W_{n,\{2,3\}}$\\
			% YMA-3
			\hline
			$(0,1,1)$   & $\{3\}$  & $ 3 $ &  $W_{n,\{0,3\}} $ &   $ W_{n,\{1,3\}} $ &   $ W_{n,\{2,3\}}$ & $/$   \\
			\hline % V=013 td = 0 —— 跨行
			$(1,0,0)$ &  $\{0,1,3\}$  &   $ 0 $ &  $/$
			&  $W_{n,\{0,1\}} \oplus W_{n,\{1,3\}}$  
			& \makecell{$W_{n,\{0,2\}} \oplus W_{n,\{1,2\}} $  \\ $ \oplus W_{n,\{2,3\}}$  }    &  $ W_{n,\{0,3\}}\oplus W_{n,\{1,3\}}$  \\
			% 新的构造部分
			\hline % V=023 td = 0 —— 跨行
			$(0,1,0)$ &  $\{0,2,3\}$  &   $ 0 $ & $/$
			&   \makecell{$W_{n,\{0,1\}} \oplus W_{n,\{1,2\}} $  \\  $ \oplus W_{n,\{1,3\}}$   } 
			& $W_{n,\{0,2\}} \oplus W_{n,\{2,3\}}$   &  $W_{n,\{0,3\}}\oplus W_{n,\{2,3\}}$\\
			\hline % V=123 td = 1 —— 跨行
			$(1,0,1)$ &  $\{1,2,3\}$  & $ 1 $ &\makecell{$W_{n,\{0,1\}} \oplus W_{n,\{0,2\}}$  \\ $  \oplus W_{n,\{0,3\}}$ }  
			& $/$
			& $W_{n,\{1,2\}} \oplus W_{n,\{2,3\}}$   &  $W_{n,\{1,3\}}\oplus W_{n,\{2,3\}}$\\
			\hline
		\end{tabular}
	\end{center}
	\vspace{-0.6cm}
\end{table*}
	\emph{Decoding}.
	To provide the decoding process for user $k$, we first define the function $g(t), t \in [0:3] $ as follows 
	\vspace{-1mm}
	\begin{align*} 
	g(0) & = (0,0,0), \quad  g(1)   = (1,1,1),  \\
	g(2) & = (0,0,1), \quad  g(3)    = (0,1,1).
	\end{align*} 
	\vspace{-1mm}
	For the sake of consistency in decoding,  for $t \in [0:3]$, we set $X^n_{g(t), \{t\}} $ equals to a zero vector with length $\frac{F}{6}$.  
	Furthermore, we set $X^n_{(1,1,0), \{0\}} = W_{n,{\{0,1\}}} \oplus W_{n,{\{0,2\}}}$, 
	$X^n_{(1,0,0), \{0\}} = W_{n,{\{0,1\}}} \oplus W_{n,{\{0,3\}}}$, 
	$X^n_{(0,1,0), \{0\}} = W_{n,{\{0,2\}}} \oplus W_{n,{\{0,3\}}}$ and 
	$X^n_{(1,0,1), \{1\}} = W_{n,{\{1,2\}}} \oplus W_{n,{\{1,3\}}}$.
	Thus, for each user $k \in [0:2]$, the decoding steps for the requested subfile $W_{D_k,  \mathcal{R} }$, $ \mathcal{R} \subset [0:3], |\mathcal{R}| = 2$,  can be expressed as follows 
	\begin{align*} 
	W_{D_k, \mathcal{R}}  
	=  \bigoplus\limits_{t \in \mathcal{V}_{\boldsymbol{d}} \setminus \mathcal{V}_{\boldsymbol{d}}\cap\mathcal{R}  } 
	Y^{(k,S_k)}_{ \{t\}\cup\mathcal{R}}   
	\oplus   \bigoplus\limits_{t \in  \mathcal{R}}
	X_{\boldsymbol{d},\mathcal{R}\setminus \{t\}} ^{g(t)(k) \oplus S_k},
	\end{align*} 
	where recall that for vector $g(t)$, $g(t)(k)$ denotes the $k$-th element of $g(t)$.	
	
	Taking $\boldsymbol{S} = (0,0,1)$, $\boldsymbol{D}=(0,0,1)$, and $\boldsymbol{D}=(0,1,1)$ as examples, we present the decoding process for user $1$. 
	For the case $\boldsymbol{D}=(0,0,1)$, we have $ \boldsymbol{d} = (0,0,0) $, user $1$ obtains $W_{0,\{0,1\}}, W_{0,\{0,2\}}, W_{0,\{0,3\}}$ from  delivery signal. 
	Adding $Y_{\{0,1,2\}}^{(1,0)} = W_{0,\{1,2\}} \oplus W_{1,\{0,2\}}\oplus W_{0,\{0,1\}}$, $X_{\boldsymbol{d},\mathcal{R} \setminus \{1\}} ^{g(1)(1)} = X_{\boldsymbol{d}, \{2\}} ^{1}  = W_{1,\{0,2\}}$ and
	$X_{\boldsymbol{d},\mathcal{R}\setminus \{2\}} ^{g(2)(1)} = X_{\boldsymbol{d},  \{1\}} ^{0} =  W_{0,\{0,1\}}$ all together, user $1$ obtains $W_{0,\{1,2\}}$.
	Similarly,  user $1$ obtains $W_{0,\{1,3\}}$ by adding $Y_{\{0,1,3\}}^{(1,0)}$, $ X_{\boldsymbol{d}, \{3\}} ^{1}$ and $ X_{\boldsymbol{d},  \{1\}} ^{1}$ 
	and obtains $W_{0,\{2,3\}}$ by adding $Y_{\{0,2,3\}}^{(1,0)}$, $ X_{\boldsymbol{d}, \{3\}} ^{0}$ and $ X_{\boldsymbol{d},  \{2\}} ^{1}$.	
	Next, consider the case of  $\boldsymbol{D}=(0,1,1)$, where  $ \boldsymbol{d} = (0,1,0) $.
	User $1$ first adds $X_{\boldsymbol{d},\{2\}}^{n} = W_{n,\{0,2\}} \oplus W_{n,\{2,3\}}$  and $X_{\boldsymbol{d},\{3\}}^{n} = W_{n,\{0,3\}}\oplus W_{n,\{2,3\}}$ to obtain $X_{\boldsymbol{d},\{0\}}^{n} = W_{n,\{0,2\}} \oplus W_{n,\{0,3\}}$, where $n=0,1$.
	For $\mathcal{R} = \{0,1\}$, 
	adding  $Y_{\{0,1,2\}}^{(1,0)} = W_{0,\{1,2\}} \oplus W_{1,\{0,2\}}\oplus W_{0,\{0,1\}}$, $Y_{\{0,1,3\}}^{(1,0)} = W_{0,\{1,3\}}\oplus W_{1,\{0,3\}} \oplus W_{1,\{0,1\}} $, $ X_{\boldsymbol{d},  \{1\}} ^{0} = W_{0,\{0,1\}} \oplus W_{0,\{1,2\}}  \oplus W_{0,\{1,3\}} $ and  	$ X_{\boldsymbol{d},  \{0\}} ^{1} = W_{1,\{0,2\}} \oplus W_{1,\{0,3\}} $  all together, user $1$ obtains $W_{1,\{0,1\}}$.   For other $\mathcal{R} $ satisfying $\mathcal{R}\subset [0:3], |\mathcal{R} |=2$, the decoding of $W_{1,{\mathcal{R} }}$ follows similar decoding steps.

	\emph{Privacy.}
	From the above example, we notice that the delivery signal only depends on $\boldsymbol{d}$. Since $\boldsymbol{D}$ and $\boldsymbol{S}$ are both i.i.d. uniform random variables, we have that $\boldsymbol{d}$ is independent of $\boldsymbol{D}$, which ensures user's privacy.

	\subsection{General Proof} \label{pfachgen}
	Now we present a general achievable scheme for the $(N,K)$ demand private coded caching problem.

	\emph{Placement Phase.}
		Split each file $W_n$ into $ \binom{NK-K+1}{r} $ nonoverlapping subfiles of equal size, i.e.,
		\begin{align*}
		W_n = \left( W_{n,\mathcal{R}} \right)_{\mathcal{R}\subset 	\mathcal{T},|\mathcal{R}| = r }, \quad  \mathcal{T} = [0:NK-K].
		\end{align*}
		The server generates a $1 \times K$  random vector $\boldsymbol{S} =(  S_0, S_1, \dots, S_{K-1})$ on a finite field $\mathbb{F}_N^{1 \times K}$, where each element in the vector is i.i.d. and uniformly distributed.
		Then for each $k$ and $S_k$, define a vector as follows
		\begin{align*} 
		  \boldsymbol{u}^{(k,S_k)} & =  \left( u_0^{(k,S_k)}, u_1^{(k,S_k)}, \dots, u_{NK-K}^{(k,S_k)} \right), \quad \text{where} \\
			 u_i^{(k,S_k)} 	 &  =   \left\{
			 \begin{aligned} 
			 & S_k  \oplus_N i, \quad \quad  i\in [0:N-1]  ,  \\
			 &  S_k  \oplus_N  \left( (i-1) \mod(N-1)\right)  ,   \\  
			 & \quad  \quad  i \in [N:(K-k)(N-1)].  \\
			 & S_k  \oplus_N  \left( (i-1)\mod(N-1)+1\right), \\
			 &  \quad  \quad  i\in [(K-k)(N-1)+1:KN-K].  
			 \end{aligned}
			 \right. 
			\end{align*}    
			\begin{remark}
				The process of generating $\boldsymbol{u}^{(k,S_k)}$ based on $k$ and $S_k$ can be seen as consisting of the following two steps: first, generate an intermediate vector $\boldsymbol{U}^{(k,S_k)} = (S_k \boldsymbol{1}_{K-k} ,( S_k \oplus_N 1)\boldsymbol{1}_{k})$; second, for this $\boldsymbol{U}^{(k,S_k)} = (\boldsymbol{U}^{(k,S_k)}_0,\dots,\boldsymbol{U}^{(k,S_k)}_{K-1})$,  
				% For a certain $\boldsymbol{d} = (d_0,\dots,d_{K-1})$, 
				let $\overline{ \boldsymbol{U}^{(k,S_k)}_0} = (\boldsymbol{U}^{(k,S_k)}_i,\boldsymbol{U}^{(k,S_k)}_i \oplus_N 1, \dots,\boldsymbol{U}^{(k,S_k)}_i \oplus_N (N-1)) $,  $\overline{\boldsymbol{U}^{(k,S_k)}_i}= (\boldsymbol{U}^{(k,S_k)}_i,\boldsymbol{U}^{(k,S_k)}_i \oplus_N 1, \dots,\boldsymbol{U}^{(k,S_k)}_i \oplus_N (N-2)) , i\in [1:K-1]$, and $ \boldsymbol{u}^{(k,S_k)}_i = (\overline{\boldsymbol{U}^{(k,S_k)}_{0}},\overline{\boldsymbol{U}^{(k,S_k)}_{1}},\dots, \overline{\boldsymbol{U}^{(k,S_k)}_{K-1}}  )  =  \left( u_0^{(k,S_k)}, u_1^{(k,S_k)}, \dots, u_{NK-K}^{(k,S_k)} \right) $.
			\end{remark} 
	The cache content  consists of two parts, $Z_{k,S_k}$ and  $S_k$, i.e.,
	\begin{align*}
	Z_{k} = &  (Z_{k,S_k},S_k),  
	\end{align*} 
	where $ Z_{k,S_k}$ is given by the delivery signal of the YMA scheme under demand $\boldsymbol{u}^{(k,S_k)}$, i.e.,  
	$X_{ \boldsymbol{u}^{(k,S_k)} }^{\text{YMA}} $. 
	For $ \forall k\in[0:K-1],S_k \in[0:N-1]$ we have  		
	% 和YMA一致的定义
	\begin{align} \label{cache}
	Y^{(k,S_k)}_{\mathcal{R}^+} 
	= & \bigoplus_{i \in \mathcal{R}^+ } W_{u^{(k,S_k)}_{i},\mathcal{R}^+ \setminus \{i\}}, \quad \mathcal{R^+} \subset \mathcal{T},|\mathcal{R^+} | = r+1, \nonumber \\
	Z_{k,S_k} = &  X_{\boldsymbol{u}^{(k,S_k)}}^{\text{YMA}} =  \{Y^{(k,S_k)}_{\mathcal{R}^+ }|\mathcal{R}^+ \cap \mathcal{K}_0 \neq \emptyset \},  
	\end{align} 
	where $\mathcal{U}_0 = [0:N-1]$ corresponds to the set of leading users in the YMA scheme. 
	The size of each $Y^{(k,S_k)}_{\mathcal{R}^+} $ is $F/{\tbinom{NK-K+1}{r}}$, thus, we have
	\begin{align*}  
	MF = \frac{\tbinom{NK-K+1}{r+1}-\tbinom{NK-K-N+1}{r+1}}{\tbinom{NK-K+1}{r}}F.
	\end{align*}

	\emph{Delivery Phase.}
	The server constructs an auxiliary  vector $\boldsymbol{d} = (d_0, d_1, \dots, d_{K-1})$ as follows
	\begin{align*}
		d_k  =  D_{k} \ominus_N S_{k}, \quad k   \in [0:K-1], 
		% 在这里general的情况是$\ominus$
	\end{align*}
	where $ \ominus_N  $ denotes  the integer subtraction modulo $N$. 

	The delivery signal consists of two parts, $X_{\boldsymbol{d}}$ and  $\boldsymbol{d} $, i.e.,
	\begin{align*}
		% 标记参数 a, i
		X_{\boldsymbol{D}} & :=  ( X_{\boldsymbol{d}},\boldsymbol{d}). 
	\end{align*} 
	To design $X_{\boldsymbol{d}}$,
	we denote the set of possible values for $\boldsymbol{d}$ by $\mathcal{D} = [0:N-1]^K $ and provide the following partition and divide it into  $\mathcal{D}_0 $, $\mathcal{D}_1 $ and $ \mathcal{D}_2$, where  
	\begin{align*}
		% 标记参数 a, i
		\mathcal{D}_0 & :=  \{(N-1) \boldsymbol{1}_{K} \}  \\ & \quad \quad    \cup  \left\{(a\boldsymbol{1}_{K-k} ,(a\oplus_N 1)\boldsymbol{1}_{k})  \right\}_{k\in[0:K-1],a\in[0:N-2]}, \nonumber \\
		\mathcal{D}_1 & :=    \left\{ a \mathbf{e}_{k} \right\}_{a\in[1:N-1],k\in [0:K-1]} , \nonumber \\
		\mathcal{D}_2 & :=  \mathcal{D} \setminus (\mathcal{D}_0 \cup \mathcal{D}_1) = [0:N-1]^K \setminus (\mathcal{D}_0 \cup \mathcal{D}_1).
	\end{align*} 
	
		To describe $X_{\boldsymbol{d}}$, where $\boldsymbol{d}\in    \mathcal{D}_{0} $,
we label the $NK-K+1$ demands in set $\mathcal{D}_{0}$ in order as $0,1,\dots,NK-K$,
the one-to-one mapping between labels and demands is defined as follows
\begin{align*}
	f: \mathcal{D}_{0}  \longmapsto \mathcal{T}, g: \mathcal{T}   \longmapsto \mathcal{D}_{0} ,
\end{align*}
where 
\begin{align*}
	f(\boldsymbol{d}) =   \left\{
	\begin{aligned}
		& f(a\boldsymbol{1}_{K})  =  a ,  \quad  \quad \quad    \quad \quad  \quad \quad a\in[0:N-1], \\
		& f((a\boldsymbol{1}_{K-k} ,(a \oplus_N  1 ) \boldsymbol{1}_{k} ) )  =  (N-1)k+a+1  ,\\   
		&  \quad  \quad \quad \quad \quad  \quad  k\in[1:K-1],  a\in[0:N-2],
	\end{aligned}
	\right. 
\end{align*}
and $g$ is the inverse function of $f$. One part of the delivery signal, i.e., $X_{\boldsymbol{d}}$, where $\boldsymbol{d} \in  \mathcal{D}_{0} $, is set as
\begin{align} \label{achxd0}
	X_{\boldsymbol{d}} 
	% =  & Z_{f(\boldsymbol{d})}^{\text{YMA}}  \nonumber \\ 
	=  	& \left( W_{n,\mathcal{R}} | f(\boldsymbol{d}) \in \mathcal{R}, \mathcal{R}\subset 	\mathcal{T} , |\mathcal{R}| = r \right)_{n\in[0:N-1]} .
\end{align}  % 删掉YMA联系部分	

To design $X_{\boldsymbol{d}}$, where $\boldsymbol{d}\in \mathcal{D} \setminus  \mathcal{D}_{0} $, 
we first define an $(NK-K+1)$-length binary vector $V_{\boldsymbol{d}}$ for each $\boldsymbol{d} \in \mathcal{D}$ as follows.
\begin{comment}
		To design $X_{\boldsymbol{d}}$, where $\boldsymbol{d}\in \mathcal{D} \setminus  \mathcal{D}_{0} $, 
	we first design a set $\mathcal{V}_{\boldsymbol{d}} \subset  \mathcal{T}$  for each demand $\boldsymbol{d} \in \mathcal{D}$, and then generate  $X_{\boldsymbol{d}}$ based on the set $\mathcal{V}_{\boldsymbol{d}}$. 
	For convenience of presentation, we define an $(NK-K+1)$-length binary vector $V_{\boldsymbol{d}}$ for each $\mathcal{V}_{\boldsymbol{d}}$. 
	The $j$-th component of $V_{\boldsymbol{d}}$ represents whether element $j$ exists in the set $\mathcal{V}_{\boldsymbol{d}} $. More specifically, $V_{\boldsymbol{d}}(j) = 1 $ means $ j \in \mathcal{V}_{\boldsymbol{d}} $, and $V_{\boldsymbol{d}}(j) = 0 $ means $ j \notin \mathcal{V}_{\boldsymbol{d}} $. The design of the set $\mathcal{V}_{\boldsymbol{d}}$ is given by the corresponding vector $V_{\boldsymbol{d}}$, which is designed as follows. 
\end{comment}
$V_{\boldsymbol{0}_K} = \mathbf{e}_{0} $  and for $\boldsymbol{d} = a \mathbf{e}_{k} \in \mathcal{D}_1$, 
\begin{align}\label{achVd2} 
	V_{\boldsymbol{d}} =   \left\{
	\begin{aligned} 
		& \mathbf{e}_{0} \oplus \bigoplus_{b\in[1:a]} \left( \mathbf{e}_{b} \oplus \mathbf{e}_{(N-1)(K-1)+b}  \right) , \quad  k=0,  \\
		& \mathbf{e}_{0} \oplus \bigoplus_{b\in[1:a]} \left( \mathbf{e}_{(N-1)(K-k)+b} \oplus \mathbf{e}_{(N-1)(K-k-1)+b}  \right) , \\  
		& \quad \quad \quad \quad \quad \quad \quad \quad \quad \quad \quad \quad \quad \quad \quad  k \in [1:K-2],  \\ 
		&  \mathbf{e}_{0} \bigoplus_{b\in[1:a]} \left( \mathbf{e}_{(N-1)+b} \oplus \mathbf{e}_{b-1}  \right) ,\quad  \quad \quad k=K-1.   
	\end{aligned}
	\right. 
\end{align}
For $\boldsymbol{d} = (d_0,d_1,\dots,d_{K-1}) \in \mathcal{D} \setminus (\mathcal{D}_1 \cup \{\boldsymbol{0}_K\}) $,  
\begin{align} 
	V_{\boldsymbol{d}} &  = \mathbf{e}_{0}  \oplus \bigoplus_{i \in[0:K-1]} \left( V_{d_i \mathbf{e}_{i}} \oplus \mathbf{e}_{0}\right).\label{achVd3}
	% 代入前面公式化简一下 ？
\end{align} 
Next, we define the set  $\mathcal{V}_{\boldsymbol{d}} \subset  \mathcal{T}$   corresponding to each $V_{\boldsymbol{d}}$, where the $j$-th component of $V_{\boldsymbol{d}}$ represents whether element $j$ exists in the set $V_{\boldsymbol{d}}$, i.e., 
\begin{align*}
\mathcal{V}_{\boldsymbol{d}} = \{ j| V_{\boldsymbol{d}}(j) = 1\}.
\end{align*}
Thus, for $\mathcal{S} \subset \mathcal{T} $ such that $ |\mathcal{S}|= r - 1$, we define 
\begin{align} \label{Xsub1}
	X^n_{\boldsymbol{d},\mathcal{S}} :=&   \bigoplus_{v\in \mathcal{V}_{\boldsymbol{d}} \setminus  \mathcal{V}_{\boldsymbol{d}} \cap \mathcal{S}  } W_{n,\{v\} \cup \mathcal{S}} , \quad \forall n\in[0:N-1].
\end{align}
For each $\boldsymbol{d} \in  \mathcal{D} \setminus \mathcal{D}_0$, we choose an element $t_{\boldsymbol{d}}$ randomly in set $\mathcal{V}_{\boldsymbol{d}}$, and then $	X_{\boldsymbol{d}}$  is set as  
		\begin{align} \label{X1}
		X_{\boldsymbol{d}} = &  \left(X^n_{\boldsymbol{d},\mathcal{S}}|
		\mathcal{S} \subset  \mathcal{T} \setminus \{t_{\boldsymbol{d}}  \} 
		,| \mathcal{S} | = r-1\right)_{n\in [0:N-1]}, % 注意一下这里是不是可以取集合相等或者说需不需要
		\end{align}
		where $\mathcal{V}_{\boldsymbol{d}} \setminus \mathcal{V}_{\boldsymbol{d}} \cap \mathcal{S} \neq \emptyset $ since $t_{\boldsymbol{d}} \in  \mathcal{V}_{\boldsymbol{d}} \setminus \mathcal{V}_{\boldsymbol{d}} \cap \mathcal{S}$.
		
		The size of each $X^j_{\boldsymbol{d},\mathcal{S}} $ is $F/{\tbinom{NK-K+1}{r}}$ and the size of $\boldsymbol{d} $ can be neglected, thus, we have
		\begin{align*}  
		RF  =     \frac{NF\tbinom{NK-K}{r-1}}{\tbinom{NK-K+1}{r}}  
		=  \frac{NFr}{NK-K+1}. 
		\end{align*}  

	\begin{remark} \label{remark4}
	 	Substituting \eqref{achVd2} into \eqref{achVd3},  
	 	for $\forall \boldsymbol{d}  \in \mathcal{D}_{0}$, we have 
	 	$V_{\boldsymbol{d}} =  \mathbf{e}_{f(\boldsymbol{d})} $ and $	\mathcal{V}_{\boldsymbol{d}} =  \{ f(\boldsymbol{d})\} $. 
	    Thus, the design of $X_{\boldsymbol{d}}$ in $\eqref{achxd0}$ also satisfies \eqref{Xsub1} and \eqref{X1}.
	    Since \eqref{Xsub1} and \eqref{X1} are the design of $X_{\boldsymbol{d}}$ for the case where $\boldsymbol{d} \in \mathcal{D} \setminus \mathcal{D}_0$, the design of $X_{\boldsymbol{d}}$ is consistent with \eqref{Xsub1} and \eqref{X1}.
	 \end{remark} 
	 
		\emph{Decoding.}
		The user $k$, $k\in[0:K-1]$, obtains its demand file $W_{D_k} = W_{d_k \oplus_N S_k}$ by computing 
		\begin{align}\label{achVd1} 
			W_{D_k , \mathcal{R}} 
			=  \bigoplus\limits_{t \in \mathcal{V}_{\boldsymbol{d}} \setminus\mathcal{V}_{\boldsymbol{d}}\cap\mathcal{R}  } 
			Y^{(k,S_k)}_{ \{t\}\cup\mathcal{R}}   
			\oplus   \bigoplus\limits_{t \in  \mathcal{R}}
			X_{\boldsymbol{d},\mathcal{R}\setminus \{t\}} ^{g(t)(k)\oplus_N S_k},
		\end{align}    
		for all $  \mathcal{R}$ satisfying  $  \mathcal{R} \subset \mathcal{T}$ and  $  |\mathcal{R}| = r $. 
		The decoding correctness is shown in Appendix  \ref{pfLemdecode}  where we prove that \eqref{achVd1} is correct and all terms in right-hand side of \eqref{achVd1}  can be obtained from $Z_{k,S_k}$ and $X_{\boldsymbol{d}}$.
		% follows from the following lemma.
		
	    \emph{Privacy.} The privacy of this scheme can be obtained through steps similar to the privacy proof in \cite[Theorem 1]{Chinmay2022}. For completeness, a detailed proof is provided in Appendix \ref{pfPrivacy}.

		Thus, the proof of Theorem \ref{ach2} is complete.
		
		\section{Proof of Theorem \ref{con2}} \label{seccon}
		In the section, to prove our converse,  we use the submodularity of the entropy function \cite{book}, i.e., let $ \mathcal{X}$  represent a set of random variables. Then for any set $\mathcal{X}_1, \mathcal{X}_2 \subset  \mathcal{X}$, we have
		\begin{align} \label{submo}
		H(\mathcal{X}_1) + H(\mathcal{X}_2) \ge H(\mathcal{X}_1 \cup \mathcal{X}_2 ) + H(\mathcal{X}_1 \cap \mathcal{X}_2 ).
		\end{align}
		Before giving the proof of Theorem \ref{con2}, we first provide the following two lemmas. 
		
		%Before giving the proof of Theorem \ref{con2},  
		%we use the submodularity of the entropy function \cite{book}, i.e., let $ \mathcal{X}$  represent a set of random variables. Then for any set $\mathcal{X}_1, \mathcal{X}_2 \subset  \mathcal{X}$, we have
		% \begin{align} \label{submo} H(\mathcal{X}_1) + H(\mathcal{X}_2) \ge H(\mathcal{X}_1 \cup \mathcal{X}_2 ) + H(\mathcal{X}_1 \cap \mathcal{X}_2 ).
		% \end{align}
		%before giving the proof of Theorem \ref{con2}, 
		%we first provide the following two lemmas. 
		The first lemma shows that from the perspective of a certain user $k$, the information it can obtain, i.e., $X$, $Z_k$ and $W_{d_k}$,  is identically distributed when the demands of other users change.
		
		\begin{Lem} \label{Lemcon11} 
			% For all $d_0,d_1,\dots, d_{K-1}, d'_0,d'_1, \dots, d'_{K-1} \in [0:N-1]$ and user $k$, $k \in [0:K-1]$, any demand private coded caching scheme satisfies the following distribution equation
			For all $d_0,d_1,\dots, d_{K-1}, d'_0,d'_1, \dots, d'_{K-1} \in [0:N-1]$ and $k \in [0:K-1]$, any demand private coded caching scheme satisfies the following distribution equation
			% 同分布
			\begin{align*} % \label{Lemcon12} 
			& \left(X_{(d_0,d_1,\dots,d_{k-1},d_k,d_{k+1}, \dots, d_{K-1})},Z_{k},W_{d_k} \right)  \nonumber \\
			\sim   & \left(X_{(d_0',d'_1,\dots,d'_{k-1},d_k, d'_{k+1},\dots, d'_{K-1})},Z_{k},W_{d_k} \right).
			\end{align*}
		\end{Lem}
		\begin{IEEEproof}
			Lemma \ref{Lemcon11} extends the result of  \cite[Lemma 4]{Chinmay2022}, which works for $K=2$, to arbitrary $K$. The proof  is similar to the proof of \cite[Lemma 4]{Chinmay2022}. For completeness, we provide the proof in Appendix \ref{pfLemcon11}.
		\end{IEEEproof}
		
		The following lemma is a converse result for the traditional $(N,K)$ coded caching problem, where there is no privacy constraint. Note that a converse for the traditional coded caching problem serves as a converse for the demand private coded caching problem.
		
		% {\color{blue}Lemma \ref{Lemcon3} } 
		% The content of these entropy terms and their combination order are carefully designed to determine the maximum number 
		\begin{Lem} \label{Lemcon3}
			For all $ a \in \{0,1\}$, $k \in [2:K]$, any coded caching scheme with $N=2$ satisfies the following inequalities,
			%\begin{subequations}
			\begin{align*}
			% \label{Lemcon31}
			& \textstyle \sum_{i=K-k}^{K-1} \sum_{j=0}^{K-1-i} 
			H\left(Z_i,X_{(a\boldsymbol{1}_{K-j},\tilde{a} \boldsymbol{1}_{j})}\right) \ge H\left(W_a, X_{a\boldsymbol{1}_{K}}\right) 
			\nonumber \\ 
			& \quad \quad  \quad  +   \textstyle \sum_{i=K-k}^{K-2}(K-i-1) H(W_a,Z_i)    +  2(k-1)F.
			\end{align*}
			%\end{subequations}
		\end{Lem}
		\begin{IEEEproof}
			The proof of Lemma \ref{Lemcon3} uses the techniques of induction and recursion. The details is in Appendix \ref{pfLemcon3}. 
		\end{IEEEproof} 
		
		Lemma \ref{Lemcon3} serves as an intermediate step in the proof of converse. More specifically, it lower bounds the joint entropy of cache content and delivery signal by terms involving the joint entropy of file and delivery signal, the joint entropy of file and cache content, and the entropy of messages. 
		
		Next, we provide the proof of \eqref{con21} and \eqref{con22}, respectively. In order to prove \eqref{con21}, first, for $k \in [2:K]$, we have
		% 要修改因为要拿掉一条线
		% \subsection{Proof of \eqref{con21}}
		\begin{subequations}
			\begin{align} 
			& \textstyle \sum_{i=K-k}^{K-1}2(K-i)H\left( Z_i\right)  \nonumber \\
			&   \quad  \quad   +\textstyle \sum_{a=0}^{1} \sum_{i=K-k}^{K-1} \sum_{j=0}^{K-1-i}   H\left(X_{(a\boldsymbol{1}_{K-j},\tilde{a} \boldsymbol{1}_{j})}\right)  \nonumber \\
			%%  --------------------------------------------------
			\ge & \textstyle \sum_{a=0}^{1} \sum_{i=K-k}^{K-1} \sum_{j=0}^{K-1-i} 
			H\left( Z_i,X_{(a\boldsymbol{1}_{K-j},\tilde{a} \boldsymbol{1}_{j})}\right)  \nonumber \\
			%% ----------------------------------------------------
			%% 应用lemma
			\label{pfcon21a} 
			\ge &  4(k-1)F +   H\left(W_0, X_{\boldsymbol{0}_{K}}  \right) +   H\left(W_1, X_{\boldsymbol{1}_{K}}  \right)   \nonumber\\
			& + \textstyle \sum_{i=K-k}^{K-2}(K-i-1) \left(H(W_0,Z_i)+H(W_1,Z_i) \right)   \\
			%% ----------------------------------------------------
			%% 应用privacy
			\label{pfcon21b} 
			= & 4(k-1)F +   H\left(W_0, X_{(\boldsymbol{0}_{K-2},1,0)}  \right) +   H\left(W_1, X_{(\boldsymbol{1}_{K-2},0,1)}  \right)   \nonumber\\
			& + H(W_0,Z_{K-2})+H(W_1,Z_{K-2})  \nonumber\\
			& +  \textstyle \sum_{i=K-k}^{K-3}(K-i-1) \left(H(W_0,Z_i)+H(W_1,Z_i) \right)  \\
			%% ----------------------------------------------------
			\label{pfcon21c}		% sub
			\ge & 4(k-1)F +   H\left(W_0  \right) + H(W_0,Z_{K-2}, X_{(\boldsymbol{0}_{K-2},1,0)})   \nonumber\\
			& +   H\left(W_1 \right)  + H(W_1,Z_{K-2}, X_{(\boldsymbol{1}_{K-2},0,1)} )  \nonumber\\
			&+  \textstyle \sum_{i=K-k}^{K-3}(K-i-1) \left(H(W_0,W_1)+H(Z_i) \right)   \\
			%% ----------------------------------------------------
			%% decode
			\label{pfcon21d}
			\ge & 4(k-1)F  +  H\left(W_0\right) +  H\left(W_1\right) +  2H\left(W_0,W_1\right) \nonumber  \\
			&+ \textstyle \sum_{i=K-k}^{K-3}(K-i-1) \left(H(W_0,W_1)+H(Z_i) \right)   \\
			%% ----------------------------------------------------
			%% 整理一下
			\label{pfcon21e}
			= & k(k+3)F + \textstyle  \sum_{i=K-k}^{K-3}  (K-1-i) H(Z_i),
			\end{align}
		\end{subequations}
		where \eqref{pfcon21a} follows from Lemma \ref{Lemcon3}, 
		\eqref{pfcon21b}  follows from the  privacy constraint, more specifically, Lemma \ref{Lemcon11},
		\eqref{pfcon21c} follows from the submodularity of the entropy function, i.e., \eqref{submo},
		\eqref{pfcon21d} follows from the correctness constraint \eqref{decoding}, and
		\eqref{pfcon21e} follows from the assumption that the files are independent, uniformly distributed   and  $F$ bits each. 
		
		Subtract the common terms of $\sum_{i=K-k}^{K-3}  (K-1-i) H(Z_i)$ from both sides of \eqref{pfcon21e}, we get
		\begin{align*} 
		& \textstyle \sum_{i=K-k}^{K-1}(K-i+1)H\left( Z_i\right) +H(Z_{K-2})   \nonumber \\ 
		& \quad  \quad    +  \textstyle  \sum_{a=0}^{1}  \sum_{j=0}^{k-1} (k-j) H\left(X_{(a\boldsymbol{1}_{K-j},\tilde{a}\boldsymbol{1}_{j})}\right) \ge  k(k+3)F.
		\end{align*}
		Since $H(Z_j) \le MF$ and $H(X_{\boldsymbol{D}}) \le RF $, for $\forall j \in [0:K-1], \forall \boldsymbol{D} \in [0:N-1]^K$,  for $\forall k\in [2:K]$,  we get
		\begin{align*} 
		\frac{(k+1)(k+2)}{2}  M +  k(k+1) R   \ge k(k+3).
		\end{align*} 
		The proof of \eqref{con21} is thus complete.
		
		Finally, we provide the proof of \eqref{con22}. 
		By replacing all $Z_i$ with $X_{(a\boldsymbol{1}_{K-i},\tilde{a} \boldsymbol{1}_{i})}$ and all $X_{(a\boldsymbol{1}_{K-i},\tilde{a} \boldsymbol{1}_{i})}$ with  $Z_i$,
		the proof of  \eqref{con22} follows the same steps with the proof of \eqref{con21} except for a few steps of using privacy constraints. 
		The detailed proof can be found in Appendix \ref{pfcon222}.
		
		The proof of Theorem \ref{con2} is thus complete.
		
		% \begin{comment}
		\section{Conclusions}
		In this paper, some progress has been made towards characterizing the exact memory-rate tradeoff for the demand private coded caching problem when there are 2 files in the system. More specifically, we first present a new virtual-user-based achievable scheme for arbitrary number of users and files. Then, for the case of 2 files and arbitrary number of users, we derive new converse bounds. As a result, we obtain
		%By exchanging cache content and corresponding  transmitted message in an existing scheme and constructing transmitted message for the remaining demands without corresponding users, we present a new {\color{red}demand private} coded caching scheme.
		%Through numerical simulation, we show that our proposed scheme outperforms known {\color{red}achievable} schemes when the cache size $M$ is small {\color{red}and $N \le K$}.   
		%In the case of $N=2,K\ge 3$, we derive new {\color{red}converse results} and obtain 
		the exact memory-rate tradeoff of the demand private coded caching problem for the case of 2 files and 3 users. As for the case of 2 files and arbitrary number of users, the exact memory-rate tradeoff is characterized  for $M\in [0,\frac{2}{K}] \cup [\frac{2(K-1)}{K+1},2]$. 
		
			\appendices
	\section{Proof of Decoding Correctness}\label{pfLemdecode} 
	In this section, we prove the correctness of decoding for the scheme described in Subsection \ref{pfachgen}. 
	More specifically, we need to show that \eqref{achVd1} is correct and all terms in right-hand side of \eqref{achVd1}  can be obtained from $Z_{k,S_k}$ and $X_{\boldsymbol{d}}$.
	Before giving the proof of decoding correctness, we first provide the following two lemmas.

	The first lemma shows the relationship between the indices of the files requested by users among different demands and serves as an intermediate step in the proof of the correctness of  \eqref{achVd1}.
	% Before we provide the  proof of lemma \ref{Lemdecode}, we first give two useful lemmas.
	\vspace{0cm} 
		\begin{Lem} \label{Lemachx2}
			For $\mathcal{R} \subset \mathcal{T}, |\mathcal{R}| = r,\boldsymbol{D} \in [0:N-1]^K ,   {\boldsymbol{S}} \in [0:N-1]^K$ and the corresponding $\boldsymbol{d} = (D_0 \ominus_N S_0, D_1 \ominus_N S_1, \dots, D_{K-1} \ominus_N S_{K-1}) $,  let 
			$
				W_{ \boldsymbol{D},\mathcal{R}}  = \left(W_{ D_0,\mathcal{R}} , W_{ D_1,\mathcal{R}} , \dots, W_{ D_{K-1},\mathcal{R}}  \right) 
			$ and $
			W_{ \boldsymbol{d},\mathcal{R}}  = \left(W_{ d_0,\mathcal{R}} , W_{ d_1,\mathcal{R}} , \dots, W_{ d_{K-1},\mathcal{R}}  \right) 
			$, we have
				\begin{subequations}
			\begin{align} \label{Lemachx21}
				W_{ \boldsymbol{d},\mathcal{R}} = &  
				\bigoplus_{t \in \mathcal{V}_{\boldsymbol{d}} } W_{g(t) ,\mathcal{R}}, \quad \text{and} \\
				 \label{Lemachx22}
				W_{ \boldsymbol{D},\mathcal{R}} = &   % W_{\boldsymbol{d},\mathcal{R}} & =
				 \bigoplus_{t \in \mathcal{V}_{\boldsymbol{d}} } W_{g(t) \oplus_N \boldsymbol{S} ,\mathcal{R}}. 
			\end{align}
				\end{subequations}
		\end{Lem}

	\vspace{0cm}
	\begin{IEEEproof} The proof can be found in Appendix \ref{pfLemachx2}.
	\end{IEEEproof}
	
	The following lemma shows that,  although some symbols $X^n_{\boldsymbol{d},\mathcal{S}}$ defined in \eqref{Xsub1} are not directly delivered in the delivery signal, these symbols can still be recovered from the delivery signal.
	\vspace{0cm}
	\begin{Lem} \label{Lemachx1}
		For $ \forall t_{\boldsymbol{d}} \in \mathcal{V}_{\boldsymbol{d}} $, 
		$\forall \boldsymbol{d} \in  \mathcal{D} \setminus \mathcal{D}_{0}$, $ \forall \mathcal{S}\subset \mathcal{T}, |\mathcal{S}| = r-1$ and $ \forall n \in [0:N-1] $,  $X^n_{\boldsymbol{d},\mathcal{S}}$ can be recovered from $X_{\boldsymbol{d}} $. 
	\end{Lem}
	\begin{IEEEproof}  The proof of this lemma  is similar to the proof of \cite[Lemma 2]{Chinmay2022}. For completeness, we provide the proof in Appendix \ref{pfLemachx1}.
	\end{IEEEproof}
	
	Next, we show the decodability for the case the auxiliary  vector $\boldsymbol{d}  \in \mathcal{D}_{0}$. If $f(\boldsymbol{d}) \in \mathcal{R} $, 
	% user $k$ can easily obtain $W_{D_k,\mathcal{R} }$ as $W_{D_k,\mathcal{R} } = X_{\boldsymbol{d}, \mathcal{R} \setminus \{ f(\boldsymbol{d})\} }^{D_k} $. In this case, 
	considering that $ \mathcal{V}_{\boldsymbol{d}} = \{f(\boldsymbol{d}) \}$, $  \mathcal{V}_{\boldsymbol{d}} \setminus\mathcal{V}_{\boldsymbol{d}} \cap\mathcal{R}  = \emptyset $,  and for all $t\in \mathcal{R} \setminus \{ f(\boldsymbol{d})\} $, $ X_{\boldsymbol{d},\mathcal{R}\setminus \{t\}} ^{n}$ equals to a zero vector with length $F/{\tbinom{NK-K+1}{r}} $, we have 
 \begin{align*} 
    & \bigoplus\limits_{t \in \mathcal{V}_{\boldsymbol{d}} \setminus\mathcal{V}_{\boldsymbol{d}} \cap\mathcal{R}  } 
 	Y^{(k,S_k)}_{ \{t\}\cup\mathcal{R}}   
 	\oplus   \bigoplus\limits_{t \in  \mathcal{R}}
 	X_{\boldsymbol{d},\mathcal{R}\setminus \{t\}} ^{g(t)(k)\oplus_N S_k} \\
 	=	&  X_{\boldsymbol{d},\mathcal{R}\setminus \{f(\boldsymbol{d})\}} ^{g(f(\boldsymbol{d}))(k)\oplus_N S_k} 
 	= X_{\boldsymbol{d},\mathcal{R}\setminus \{f(\boldsymbol{d})\}} ^{d_k \oplus_N S_k} 
 	= W_{D_k , \mathcal{R}}, 
 \end{align*} 
which is  consistent with \eqref{achVd1}. If    $f(\boldsymbol{d}) \notin \mathcal{R}$, we have \begin{align} \label{D1}
	& \bigoplus\limits_{t \in \mathcal{V}_{\boldsymbol{d}} \setminus\mathcal{V}_{\boldsymbol{d}} \cap\mathcal{R}  } 
	Y^{(k,S_k)}_{ \{t\}\cup\mathcal{R}}   
	\oplus   \bigoplus\limits_{t \in  \mathcal{R}}
	X_{\boldsymbol{d},\mathcal{R}\setminus \{t\}} ^{g(t)(k)\oplus_N S_k}  \nonumber \\
 	 \overset{(a)}{=}   &  
	Y^{(k,S_k)}_{ \{f(\boldsymbol{d})\}\cup\mathcal{R}}   
	\oplus   \bigoplus\limits_{t \in  \mathcal{R}}
	X_{\boldsymbol{d},\mathcal{R}\setminus \{t\}} ^{g(t)(k)\oplus_N S_k}  \nonumber \\
	% ----------------------------------------------
	 \overset{(b)}{=}   &  
	\bigoplus_{i \in  \{f(\boldsymbol{d})\}\cup\mathcal{R} } W_{u^{(k,S_k)}_{i}, \{f(\boldsymbol{d})\}\cup\mathcal{R}  \setminus \{i\}} 
	 \nonumber \\  &  \quad \quad  \quad \quad \oplus   \bigoplus\limits_{t \in  \mathcal{R}}
	W_{g(t)(k)\oplus_N S_k, \{f(\boldsymbol{d}  )\} \cup \mathcal{R}\setminus \{t\}}   \nonumber \\
	% ------------------------------------------------
	=   &  
	W_{u^{(k,S_k)}_{f(\boldsymbol{d} )},\mathcal{R}} \oplus   
	\bigoplus_{t \in \mathcal{R} } \big( W_{u^{(k,S_k)}_{t}, \{f(\boldsymbol{d})\}\cup\mathcal{R}  \setminus \{t\}} 
	   \nonumber \\  &  \quad \quad  \quad \quad  \oplus 
	W_{g(t)(k)\oplus_N S_k, \{f(\boldsymbol{d}  )\} \cup \mathcal{R}\setminus \{t\}}  \big)  \nonumber  \\
	 \overset{(c)}{=}  & W_{D_k , \mathcal{R}},  
	\end{align}
	where $(a)$ follows from $\mathcal{V}_{\boldsymbol{d}} \setminus\mathcal{V}_{\boldsymbol{d}} \cap\mathcal{R}  =  \mathcal{V}_{\boldsymbol{d}} = f(\boldsymbol{d})$,
	$(b)$  follows from the design of the cache content and the delivery signal, i.e., \eqref{cache} and  \eqref{Xsub1}, and $(c)$ follows from  the fact that for $ \boldsymbol{d}  = (a\boldsymbol{1}_{K-i} ,(a\oplus_N 1) \boldsymbol{1}_{i})  \in  \mathcal{D}_0$,  $ u^{(k,S_k)}_{f(\boldsymbol{d})} = a \oplus_N S_k $  when $k+i<K$, $ u^{(k,S_k)}_{f(\boldsymbol{d})} = a \oplus_N 1 \oplus_N S_k $  when $k+i \ge K$ and $ u^{(k,S_k)}_{t} = g(t)(k)\oplus_N S_k $.
	Thus, for the case  $\boldsymbol{d}  \in \mathcal{D}_{0}$, \eqref{achVd1} is proved to be correct.
	
 	For convenience of presentation, let $c_t = g(t)(k)\oplus_N S_k$ denote the demand of user $k$ with key $S_k$ under demand vector $g(t)$.
 	Thus, from the decoding steps shown in  \eqref{D1},  
 for $\forall t\in \mathcal{T}$ and $\mathcal{R}$ satisfying $ t \notin \mathcal{R} , \mathcal{R} \subset \mathcal{T}, |\mathcal{R}| = r $, we have 
	\begin{align*} % \label{decoded0}
		& W_{c_t,\mathcal{R}} = Y^{(k,S_k)}_{ \{t\}\cup\mathcal{R}} \oplus  \bigoplus\limits_{u \in \mathcal{R}} W_{c_u, \{t\} \cup \mathcal{R} \setminus  \{u\} }.
	\end{align*}
	For all demands $\boldsymbol{d} \in \mathcal{D} \setminus \mathcal{D}_0$, all users $k$ with the key $ S_k$, $k\in[0:K-1], S_k\in[0:N-1]$, and all $  \mathcal{R}$  satisfying  $  \mathcal{R} \subset \mathcal{T}$ and  $  |\mathcal{R}| = r $, we have
	% 修改公式: 
	\begin{subequations}
		\begin{align} 
			& W_{D_k, \mathcal{R}}  =  W_{d_k \oplus_{N} S_k , \mathcal{R}} \nonumber \\
			% ----------------------------------------- 
			% 第一步（lemma2）
			\label{pfach_a}
			= & \bigoplus_{t \in \mathcal{V}_{\boldsymbol{d}} } W_{ {g} (t)(k) \oplus_N S_k,\mathcal{R}}  
			= \bigoplus_{t \in \mathcal{V}_{\boldsymbol{d}} }W_{c_t,\mathcal{R}} \\
			% -------------------------------------------- 
			% 第1.2步
			= & \bigoplus_{t \in \mathcal{V}_{\boldsymbol{d} }\cap \mathcal{R} }W_{c_t,\mathcal{R}} \nonumber \\
			& \quad \oplus  \bigoplus_{t \in \mathcal{V}_{\boldsymbol{d} } \setminus \mathcal{V}_{\boldsymbol{d}  } \cap \mathcal{R} }  \left\{ Y^{(k,S_k)}_{ \{t\}\cup\mathcal{R}} \oplus  \bigoplus\limits_{u \in \mathcal{R}} W_{c_u, \{t\} \cup \mathcal{R} \setminus  \{u\} } \right\}\nonumber  \\
			% ----------------------------------------------------
			% 第二步（和[12]中步骤一致）
			\label{pfach_b}
			= & \bigoplus\limits_{t \in \mathcal{V}_{\boldsymbol{d}} \setminus\mathcal{V}_{\boldsymbol{d}}\cap\mathcal{R}  } 
			Y^{(k,S_k)}_{ \{t\}\cup\mathcal{R}}  \nonumber \\
			& \oplus   \bigoplus\limits_{t \in  \mathcal{V}_{\boldsymbol{d}}\cap\mathcal{R}  }
			\left\{ \bigoplus\limits_{u \in \mathcal{V}_{\boldsymbol{d}} \setminus\mathcal{V}_{\boldsymbol{d}}\cap(\mathcal{R} \setminus \{t\})  } 
			W_{c_t,\{u\} \cup \mathcal{R}\setminus \{t\}}\right\}   \nonumber \\
			& \oplus   \bigoplus\limits_{t \in  \mathcal{R} \setminus \mathcal{V}_{\boldsymbol{d}}\cap\mathcal{R}  }
			\left\{ \bigoplus\limits_{u \in \mathcal{V}_{\boldsymbol{d}} \setminus\mathcal{V}_{\boldsymbol{d}}\cap(\mathcal{R} \setminus \{t\})  } 
			W_{c_t,\{u\} \cup \mathcal{R}\setminus \{t\}}\right\}   \\
			% ----------------------------------------------------
			% 第三步 带入一下Z X 设计
			\label{pfach_c}
			= & \bigoplus\limits_{t \in \mathcal{V}_{\boldsymbol{d}} \setminus\mathcal{V}_{\boldsymbol{d}}\cap\mathcal{R}  } 
			Y^{(k,S_k)}_{ \{t\}\cup\mathcal{R}}   
			\oplus   \bigoplus\limits_{t \in \mathcal{R} }
			X_{\boldsymbol{d},\mathcal{R}\setminus \{t\}} ^{c_t},
		\end{align}
	\end{subequations}
	where \eqref{pfach_a} follows from Lemma \ref{Lemachx2}, \eqref{pfach_b} follows  from  \cite[Equations (30) and (31)]{Chinmay2022},
	% follow 的类似步骤没有写
	and \eqref{pfach_c} follows from the design of the delivery signal, i.e., \eqref{Xsub1}. 
	Since from the YMA scheme\cite{Qian2018}, all symbols $Y^{(k,S_k)}_{\mathcal{R}^+}$  such that $\mathcal{R}^+ \subset [0:NK-K] $ and  $|\mathcal{R}^+ | =r+1 $ can be obtained from $X_{\boldsymbol{u}^{(k,S_k)}}^{\text{YMA}}$, 
	the first term of \eqref{pfach_c} can be obtained from cache content, i.e., $Z_{k,S_k}$. 
	The second term of \eqref{pfach_c} can be obtained from delivery signal, i.e., $X_{\boldsymbol{d}} $,  following from Lemma \ref{Lemachx1}. 
	
	Hence, the proof of decoding correctness is complete, provided we prove Lemmas \ref{Lemachx2} and \ref{Lemachx1}. These will be proved next.
	
	%{\color{blue}Hence, the proof of Lemma \ref{Lemdecode} is complete, provided we prove Lemmas \ref{Lemachx2} and \ref{Lemachx1}. These will be proved next. }

	\subsection{Proof of Lemma \ref{Lemachx2}} \label{pfLemachx2}
	We first prove the correctness of \eqref{Lemachx21}. 
	For $\boldsymbol{d}\in \mathcal{D}_{0} $, \eqref{Lemachx21} clearly holds.
	For $\boldsymbol{d} = a\mathbf{e}_k \in \mathcal{D}_1,a\in[1:N-1]$, when $k=0$, we have
	% \begin{subequations}
		\begin{align*}  
			&  \bigoplus_{t \in \mathcal{V}_{\boldsymbol{d}} } W_{g(t),\mathcal{R}} \nonumber \\
			\overset{\text{(a)}}{=}  
			& W_{(\boldsymbol{0}_K),\mathcal{R}}  \oplus \left( W_{0,\mathcal{R}} \oplus  W_{a,\mathcal{R}}, 0,\dots,0 \right) \nonumber \\
			=
			&  \left( W_{a,\mathcal{R}},  W_{0,\mathcal{R}},\dots, W_{0,\mathcal{R}} \right)
			=  W_{\boldsymbol{d},\mathcal{R}} ,
		\end{align*}
		% \end{subequations}
	where (a) follows from the design of $V_{\boldsymbol{d}}$, i.e., \eqref{achVd2}, 
	more specifically, user $i$, $i\in[1:N-1]$, has the same demand  $d_i=b $ under demands $g(b)$ and  $g((N-1)(K-1)+b)$ for $b\in[1:a]$, user $0$ has the same demand  under $d_i=b $ under demands $g(b)$ and  $g((N-1)(K-1)+b+1)$ for $b\in[1:a-1]$. 
	Similarly, when $k\in[1:K-1]$, we have
	\begin{align*}  
		\bigoplus_{t \in \mathcal{V}_{\boldsymbol{d}} } W_{g(t),\mathcal{R}}
		= &  W_{\boldsymbol{0}_K,\mathcal{R}}  
		\oplus \left( \boldsymbol{0}_{k}, W_{0,\mathcal{R}} \oplus  W_{a,\mathcal{R}},  \boldsymbol{0}_{K-k-1} \right) \nonumber \\
		= &  W_{a\mathbf{e}_k,\mathcal{R}} =  W_{\boldsymbol{d},\mathcal{R}} .
	\end{align*}
	% \end{subequations}
	Thus, \eqref{Lemachx21} holds for $\boldsymbol{d}\in \mathcal{D}_1 $. Finally, consider $\boldsymbol{d} \in \mathcal{D} \setminus {(\mathcal{D}_1 \cup \mathcal{D}_{0}) }$, we have  
	\begin{align*}  
	&  \bigoplus_{t \in \mathcal{V}_{\boldsymbol{d}} } W_{g(t),\mathcal{R}} \nonumber \\
	\overset{\text{(b)}}{=}  
	& W_{\boldsymbol{0}_K,\mathcal{R}}  \oplus 
	\bigoplus_{k \in[0:K-1]} \left( \boldsymbol{0}_{k}, W_{0,\mathcal{R}} \oplus  W_{d_k,\mathcal{R}},  \boldsymbol{0}_{K-k-1} \right)   \nonumber \\
	= &  W_{\boldsymbol{d},\mathcal{R}} ,
	\end{align*} 
	% \end{subequations}
where (b) follows from the design of $V_{\boldsymbol{d}}$, i.e.,  \eqref{achVd3} and the correctness of equation \eqref{Lemachx21} for $\boldsymbol{d}\in \mathcal{D}_1 \cup \mathcal{D}_{0}$. 
The proof of \eqref{Lemachx21} is thus complete. 

In the above proof, all the $W_{n,\mathcal{R}}, n\in [0:N-1]$, involved are changed to $W_{n \oplus_N S_k,\mathcal{R}}$, and the proof still holds. Therefore,  \eqref{Lemachx22} clearly holds.

The proof of Lemma \ref{Lemachx2}  is thus complete.

\subsection{Proof of Lemma \ref{Lemachx1}} \label{pfLemachx1}
When the randomly choosen $t_{\boldsymbol{d}} \notin \mathcal{S}$, $X^n_{\boldsymbol{d},\mathcal{S}}$  already exists in $X_{\boldsymbol{d}}$. 
We only need to prove  $\forall X^n_{\boldsymbol{d},\mathcal{S}},  \mathcal{S} \subset   \mathcal{T}, t_{\boldsymbol{d}}  \in \mathcal{S}  $ can be recover from $ X_{\boldsymbol{d}}$.
First we define $\mathcal{A}:= \mathcal{S} \setminus \{t_{\boldsymbol{d} } \}$, and we have
\begin{subequations}
% 直接从头证明版本
\begin{align}
% 第一步
X^n_{\boldsymbol{d},\mathcal{S}} =&   
\bigoplus_{t\in \mathcal{V}_{\boldsymbol{d}} \setminus \mathcal{V}_{\boldsymbol{d}}  \cap \mathcal{S}  } 
W_{n, \mathcal{S}  \cup \{t\} } \label{lemachxpf1}  \\
% 第二步
=&   \bigoplus_{t\in \mathcal{V}_{\boldsymbol{d}} \setminus \mathcal{V}_{\boldsymbol{d}}  \cap \mathcal{S}  } 
W_{n, \mathcal{A} \cup \{t,t_{\boldsymbol{d} }\} }   \label{lemachxpf2} \\
% 第三步
=&   \bigoplus_{t\in \mathcal{V}_{\boldsymbol{d}} \setminus \mathcal{V}_{\boldsymbol{d}}  \cap \mathcal{S}  }  
\Bigg( % 内层 1
\bigoplus_{v\in \mathcal{V}_{\boldsymbol{d}} \setminus  \mathcal{V}_{\boldsymbol{d}} \cap (\mathcal{A}\cup \{t\})   } 
W_{n,\mathcal{A}  \cup \{t,v\} } \nonumber \\ 
& \quad \quad   \quad  \quad  \quad    \oplus   % 内层 2
\bigoplus_{v \in \mathcal{V}_{\boldsymbol{d}} \setminus \mathcal{V}_{\boldsymbol{d}}  \cap (\mathcal{S}\cup \{ t\})  }  
W_{n, \mathcal{A} \cup \{t,v\} }  \Bigg)
\label{lemachxpf3} \\
% 第四步
=&   \bigoplus_{t\in \mathcal{V}_{\boldsymbol{d}} \setminus \mathcal{V}_{\boldsymbol{d}}  \cap \mathcal{S}  }  \Bigg(
X^n_{\boldsymbol{d},\mathcal{A}\cup \{ t \}} \nonumber \\
& \quad \quad   \quad  \quad  \quad    \oplus
\bigoplus_{v \in \mathcal{V}_{\boldsymbol{d}} \setminus \mathcal{V}_{\boldsymbol{d}}  \cap (\mathcal{S}\cup \{ t\})  }  
W_{n, \mathcal{A} \cup \{t,v\} }  \Bigg)
\label{lemachxpf4} \\
% —— 直接拆开
=&   \bigoplus_{t\in \mathcal{V}_{\boldsymbol{d}} \setminus \mathcal{V}_{\boldsymbol{d}}  \cap \mathcal{S}  }  
X^n_{\boldsymbol{d},\mathcal{A}\cup \{ t \}}  \nonumber \\
&  \quad   \oplus % 后一半	
\bigoplus_{t\in \mathcal{V}_{\boldsymbol{d}} \setminus \mathcal{V}_{\boldsymbol{d}}  \cap \mathcal{S}  }  
\bigoplus_{v \in \mathcal{V}_{\boldsymbol{d}} \setminus \mathcal{V}_{\boldsymbol{d}}  \cap (\mathcal{S}\cup \{ t\})  }  
W_{n, \mathcal{A} \cup \{t,v\} }  \nonumber \\
% 第五步 —— 后半=0
=&   \bigoplus_{t\in \mathcal{V}_{\boldsymbol{d}} \setminus \mathcal{V}_{\boldsymbol{d}}  \cap \mathcal{S}  }  
X^n_{\boldsymbol{d},\mathcal{A}\cup \{ t \}}, 
\label{lemachxpf5} 
\end{align}
\end{subequations}
% 第一步原因 % 第四步原因
where \eqref{lemachxpf1} and \eqref{lemachxpf4}  follow  from the definition of $	X^n_{\boldsymbol{d},\mathcal{S}} $, i.e.,  \eqref{Xsub1},
% 第二步原因
\eqref{lemachxpf2}  follows  from the fact that  $t \neq t_{\boldsymbol{d}} $,  $\forall t \in \mathcal{V}_{\boldsymbol{d} } \setminus \mathcal{V}_{\boldsymbol{d} } \cap\mathcal{S} $,
% As $ t_{\boldsymbol{d}}  \in \mathcal{S} $ and $ t_{\boldsymbol{d}}  \in \mathcal{V}_{\boldsymbol{d} }  $ , for all $ t \in \mathcal{V}_{\boldsymbol{d} } \setminus \mathcal{V}_{\boldsymbol{d} } \cap\mathcal{S} $, we have $t \neq t_{\boldsymbol{d}} $, so we can write \eqref{lemachxpf1} into \eqref{lemachxpf2}.
% 第三步原因
\eqref{lemachxpf3} follows from the fact that $ \left(  \mathcal{V}_{\boldsymbol{d}} \setminus \mathcal{V}_{\boldsymbol{d}}  \cap (\mathcal{S}\cup \{ t\}) \right)  \cup \{t_{\boldsymbol{d} } \} =  \mathcal{V}_{\boldsymbol{d}} \setminus  \mathcal{V}_{\boldsymbol{d}} \cap (\mathcal{A}\cup \{t\})  $, 
% 第五步原因
and \eqref{lemachxpf5} follows from the fact that every term $W_{n, \mathcal{A} \cup \{t,v\} }  $ appears twice in the binary addition.

The proof of Lemma \ref{Lemachx1}  is thus complete. 

\section{Proof of  Privacy}
\label{pfPrivacy}	
In this section, we present the proof of privacy of the scheme in  Theorem \ref{ach2}, namely the proof that \eqref{privacy2} holds in the proposed scheme.
For any $k\in[0:K-1]$, we have 
\begin{subequations}
	\begin{align}
	& I \left(D_{[0:K-1]\setminus \{k\}}; X_{\boldsymbol{D}} , Z_k, D_k \right)  \nonumber \\
	% -----------?
	\le   &  I \left(D_{[0:K-1]\setminus \{k\}}; X_{\boldsymbol{D}} , Z_k, D_k, W_{[0:N-1]} \right) \nonumber   \\
	% -------------------------------------
	= & I \left(D_{[0:K-1]\setminus \{k\}};  W_{[0:N-1]} \right)  \nonumber \\
	& + I \left(D_{[0:K-1]\setminus \{k\}}; X_{\boldsymbol{D}} , Z_k, D_k|  W_{[0:N-1]} \right)  \nonumber  \\
	=  	& \label{20a}
	 H \left(  X_{\boldsymbol{D}} , Z_k, D_k|  W_{[0:N-1]} \right) \nonumber \\
	&  -H \left( X_{\boldsymbol{D}} , Z_k, D_k|  W_{[0:N-1]},D_{[0:K-1]\setminus \{k\}} \right) \\
	=  \label{21b} 
	&  H \left(   \boldsymbol{d}  , S_k, D_k|  W_{[0:N-1]} \right) \nonumber \\
	&  -H \left( \boldsymbol{d}  , S_k, D_k|  W_{[0:N-1]},D_{[0:K-1]\setminus \{k\}} \right) \\
	=   \label{21c}  	&  H \left(   \boldsymbol{d}  , S_k, D_k  \right)   -H \left( \boldsymbol{d}  , S_k, D_k|   D_{[0:K-1]\setminus \{k\}} \right) \\
	= \label{21d}   & 0,
	\end{align}
\end{subequations} 
where \eqref{20a}  follows from the fact that $D_{ [0:K-1]}$ are independent of all files $W_{[0:N-1]} $, 
\eqref{21b}  follows from the  fact that  
$H(Z_{k,S_k}| W_{[0:N-1]} ) = 0$ and $H(X_{\boldsymbol{d}}| W_{[0:N-1]} ) = 0 $,
\eqref{21c}  follows from the fact that  
$( \boldsymbol{d}, S_k, D_k )$ is independent of all files $W_{[0:N-1]}$,
and \eqref{21d}  follows from the fact that  
$d_k = D_k \ominus_N S_k$ is independent of  $ D_k $. 
Due to the fact that  mutual information is greater than or equal to zero, we obtain $  I \left(D_{[0:K-1]\setminus \{k\}}; X_{\boldsymbol{D}} , Z_k, D_k \right) = 0$ and  the privacy of the scheme is thus proved.

\section{Proof of  \eqref{con22} \label{pfcon222} }
In this section, we present the proof of \eqref{con22}.
Similar to the proof of  \eqref{con21}, 
we first provide the following lemma.
\begin{Lem} \label{Lemcon3b}
For all $ a \in \{0,1\}$, $k \in [2:K]$, any coded caching scheme with $N=2$ satisfies the following inequalities,
\begin{align*}
% \label{Lemcon32}
&\sum_{i=K-k}^{K-1} \sum_{j=0}^{K-1-i}  H\left( 	X_{(a\boldsymbol{1}_{K-i},\tilde{a} \boldsymbol{1}_{i})},Z_j\right) \ge H\left(W_a, Z_{0}\right) \nonumber \\ 
& + \sum_{i=K-k}^{K-2}  (K-1-i) H(W_a,X_{(a\boldsymbol{1}_{K-i},\tilde{a} \boldsymbol{1}_{i})})   +  2(k-1)F .
\end{align*}
\end{Lem}
\begin{IEEEproof}
By replacing all $Z_i$ with $X_{(a\boldsymbol{1}_{K-i},\tilde{a} \boldsymbol{1}_{i})}$ and all $X_{(a\boldsymbol{1}_{K-i},\tilde{a} \boldsymbol{1}_{i})}$ with  $Z_i$, the proof of  Lemma \ref{Lemcon3b} follows the same steps as the proof of Lemma  \ref{Lemcon3}.
The detailed proof can be found in Appendix \ref{pfLemcon3}.
\end{IEEEproof}

Next, in order to prove \eqref{con22}, first, for $k \in [2:K]$, we have
\begin{subequations}
\begin{align} 
% \\ 合适选择Z和X
&  \textstyle  \sum_{a=0}^{1} \sum_{i=K-k}^{K-1}  (K-i) 
H\left( X_{(a\boldsymbol{1}_{K-i},\tilde{a} \boldsymbol{1}_{i})} \right)  \nonumber  \\
& \quad \quad \quad \quad \quad \quad 
+   \textstyle \sum_{i=K-k}^{K-1}  \sum_{j=0}^{K-1-i} 2H\left( Z_j\right)   \nonumber  \\
%% ----------------------------------------------------
\ge & \textstyle  \sum_{a=0}^{1} \sum_{i=K-k}^{K-1} \sum_{j=0}^{K-1-i}
H\left( X_{(a\boldsymbol{1}_{K-i},\tilde{a} \boldsymbol{1}_{i})},Z_j  \right)  \nonumber   \\
%% ----------------------------------------------------
%% 应用递推
\label{pfcon22a}
\ge & 4(k-1)F +  H\left(W_0, Z_{0}\right) +  H\left(W_1, Z_{0}\right)  \nonumber \\
& +  \textstyle \sum_{i=K-k}^{K-2}  (K-1-i) \big(H(W_0,X_{(\boldsymbol{0}_{K-i},\boldsymbol{1}_{i})}) \nonumber \\  
& \quad \quad \quad \quad \quad  \quad  \quad  \quad  \quad \quad \quad + H(W_1,X_{(\boldsymbol{1}_{K-i},\boldsymbol{0}_{i})})  \big)  \\
%% ----------------------------------------------------
%% 整理一下(可注释掉)
= & 4(k-1)F  +  H\left(W_0, Z_{0}\right) +  H\left(W_1, Z_{0}\right)  \nonumber \\
& +    H(W_0,X_{(\boldsymbol{0}_{2},\boldsymbol{1}_{K-2})}) +  H(W_1,X_{(\boldsymbol{1}_{2},\boldsymbol{0}_{K-2})})  \nonumber  \\
& +   (k-1) \left(H(W_0,X_{(\boldsymbol{0}_{k},\boldsymbol{1}_{K-k})}) +  H(W_1,X_{(\boldsymbol{1}_{k},\boldsymbol{0}_{K-k})})  \right) \nonumber  \\
& +  \textstyle \sum_{i=K-k+1}^{K-3}  (K-1-i) \big(H(W_0,X_{(\boldsymbol{0}_{K-i},\boldsymbol{1}_{i})}) \nonumber \\  
& \quad \quad \quad \quad \quad  \quad  \quad  \quad \quad \quad  \quad  + H(W_1,X_{(\boldsymbol{1}_{K-i},\boldsymbol{0}_{i})})  \big)  \nonumber   \\
%% ----------------------------------------------------
%% 应用privacy
\label{pfcon22b}
= &  4(k-1)F  +  H\left(W_0, Z_{0}\right) +  H\left(W_1, Z_{0}\right)  \nonumber \\
& +    H(W_0,X_{(1,0,\boldsymbol{1}_{K-2})}) +  H(W_1,X_{(0,1,\boldsymbol{0}_{K-2})})  \nonumber  \\
& +   (k-1) \left(H(W_0,X_{(\boldsymbol{0}_{K-1},1)}) +  H(W_1,X_{(\boldsymbol{0}_{K-1},1)})  \right) \nonumber  \\
& + \textstyle  \sum_{i=K-k+1}^{K-3}  (K-1-i) \big(H(W_0,X_{(\boldsymbol{0}_{K-i},\boldsymbol{1}_{i})}) \nonumber \\  
& \quad \quad \quad \quad \quad  \quad  \quad  \quad  \quad  \quad  \quad  + H(W_1,X_{(\boldsymbol{0}_{K-i},\boldsymbol{1}_{i})})  \big)     \\
%% ----------------------------------------------------
%% sub
\label{pfcon22c}
\ge & 4(k-1)F  +  H\left(W_0\right) +  H\left(W_1\right)  \nonumber \\
& +    H(W_0,Z_0,X_{(1,0,\boldsymbol{1}_{K-2})}) +  H(W_1,Z_0,X_{(0,1,\boldsymbol{0}_{K-2})})  \nonumber  \\
& +   (k-1) \left(H(W_0,W_1) +  H(X_{(\boldsymbol{0}_{K-1},1)})  \right) \nonumber  \\
& +  \textstyle \sum_{i=K-k+1}^{K-3}  (K-1-i) \big(H(W_0,W_1)  \nonumber \\  
& \quad \quad \quad \quad \quad  \quad  \quad  \quad  \quad  \quad  \quad  + H(X_{(\boldsymbol{0}_{K-i},\boldsymbol{1}_{i})})  \big) \\
%% ----------------------------------------------------
%% decode, privacy
\label{pfcon22d}
\ge & 4(k-1)F  +  H\left(W_0\right) +  H\left(W_1\right) +  2H\left(W_0,W_1\right) \nonumber  \\
& +   (k-1) \left(H(W_0,W_1) +  H(X_{(\boldsymbol{0}_{k},\boldsymbol{1}_{K-k})})  \right) \nonumber    \\
& +  \textstyle \sum_{i=K-k+1}^{K-3}  (K-1-i) \big(H(W_0,W_1)  \nonumber \\  
& \quad \quad \quad \quad \quad  \quad  \quad  \quad  \quad  \quad  \quad  + H(X_{(\boldsymbol{0}_{K-i},\boldsymbol{1}_{i})})  \big) \\
%% ----------------------------------------------------
%% 整理一下
\label{pfcon22e}
= & k(k+3)F + \textstyle \sum_{i=K-k}^{K-3}  (K-1-i) H \left( X_{(\boldsymbol{0}_{K-i},\boldsymbol{1}_{i})}  \right)    
\end{align}
\end{subequations}
where \eqref{pfcon22a} follows from Lemma \ref{Lemcon3b}, 
\eqref{pfcon22b} and the term $ H(X_{(\boldsymbol{0}_{k},\boldsymbol{1}_{K-k})}) $ in \eqref{pfcon22d} follow  from the privacy constraint, more specifically, Lemma \ref{Lemcon11},
\eqref{pfcon22c} follows from the submodularity of the entropy function, i.e., \eqref{submo},
the term $2H(W_0,W_1)$ in \eqref{pfcon22d} follows from the correctness constraint \eqref{decoding},
and \eqref{pfcon22e} follows from the assumption that the files are independent, uniformly distributed  and  $F$ bits each. 

Subtract the common terms of $\sum_{i=K-k}^{K-3}  (K-1-i) H \left( X_{(\boldsymbol{0}_{K-i},\boldsymbol{1}_{i})}  \right)   $ from both sides of \eqref{pfcon22e}, we get
\begin{align*} 
% \\ 合适选择Z和X
&   \textstyle \sum_{i=K-k}^{K-1} \left( (K-i) 
H\left( X_{(\boldsymbol{1}_{K-i},\boldsymbol{0}_{i})} \right) + H\left( X_{(\boldsymbol{0}_{K-i},\boldsymbol{1}_{i})} \right)   \right) \\
&  +  H\left( X_{(\boldsymbol{0}_{2},\boldsymbol{1}_{K-2})} \right)  + \textstyle \sum_{j=0}^{k-1} 2 (k-j)  H\left( Z_j\right)      \ge k(k+3)F.
\end{align*}
Since $H(Z_j) \le MF$ and $H(X_{\boldsymbol{D}}) \le RF $, for   $\forall j \in [0:K-1], \forall \boldsymbol{D} \in [0:N-1]^K$, we have
\begin{align*} 
% \\ 合适选择Z和X
k(k+1) M +  \frac{(k+1)(k+2)}{2} R   \ge k(k+3).
\end{align*}
The proof of \eqref{con22}  is thus complete.

\section{Proof of Lemma \ref{Lemcon11}} \label{pfLemcon11} 
% e prove \eqref{Lemcon12} for $k=0$ and 
Since any demand private coded caching scheme satisfies  $I \left(D_{[0:K-1]\setminus \{k\}};X,Z_{k} ,D_{k} \right) = 0 $ and $H ( W_{D_k}|X, Z_k,D_k) = 0 $, 
we have  $ I \left( D_{[0:K-1]\setminus \{k\}} ; X, Z_k,D_k , W_{D_k}\right) =0$.
Then we have 
\begin{align*}
& 	\text{Pr} \Big(D_{[0:K-1] \setminus \{k\} } = (d_0, \dots,d_{k-1}, d_{k+1}\dots  d_{K-1})   \nonumber \\
& \quad	\quad \quad \quad \quad \quad | X=x, Z_k = z_k, W_{d_k} = w_{d_k},D_k = d_k \Big) \nonumber \\
= & \text{Pr} \Big(D_{[0:K-1] \setminus \{k\} } = (d_0', \dots,d'_{k-1}, d'_{k+1}\dots  d'_{K-1})   \nonumber \\
& \quad	\quad \quad \quad \quad \quad | X=x, Z_k = z_k, W_{d_k} = w_{d_k},D_k = d_k \Big) .
\end{align*}
Multiply both sides of the above equation by the following equation simultaneously
\begin{align*}
\text{Pr} \big( X=x, Z_k = z_k, W_{d_k} = w_{d_k}|D_k = d_k \big),
\end{align*}
we have
\begin{align} \label{Pr1}
& 	\text{Pr} \Big(D_{[0:K-1] \setminus \{k\} } = (d_0, \dots,d_{k-1}, d_{k+1}\dots  d_{K-1}) ,  \nonumber \\
& \quad	\quad \quad \quad     X=x, Z_k = z_k, W_{d_k} = w_{d_k}| D_k = d_k \Big) \nonumber \\
= & \text{Pr} \Big(D_{[0:K-1] \setminus \{k\} } = (d_0', \dots,d'_{k-1}, d'_{k+1}\dots  d'_{K-1}),   \nonumber \\
& \quad	\quad \quad \quad    X=x, Z_k = z_k, W_{d_k} = w_{d_k} | D_k = d_k \Big) .
\end{align}
Since $D_{[0:K-1]}$ are all i.i.d. random variables and all uniformly distributed over $\{0,1,\dots ,N-1\} $, we have
\begin{align}\label{Pr2}
& 	\text{Pr} \big(D_{[0:K-1] \setminus \{k\} } = (d_0, \dots,d_{k-1}, d_{k+1}\dots  d_{K-1})| D_k = d_k \big) \nonumber \\
= & \text{Pr} \big(D_{[0:K-1] \setminus \{k\} } = (d_0', \dots,d'_{k-1}, d'_{k+1}\dots  d'_{K-1})| D_k = d_k \big) . 
\end{align}
We have that the ratio of the left-hand side of \eqref{Pr1} and  \eqref{Pr2} is equal to the ratio of the right-hand side of \eqref{Pr1} and  \eqref{Pr2}, i.e.,
\begin{align*} 
& 	\text{Pr} \Big( X=x, Z_k = z_k, W_{d_k} = w_{d_k}  \nonumber \\
& \quad	\quad \quad \quad   \quad  \quad  \quad  \quad  \quad   \big| D_{[0:K-1] } = (d_0,  d_{1}\dots  d_{K-1}) \Big) \nonumber \\
= & \text{Pr} \Big(X=x, Z_k = z_k, W_{d_k} = w_{d_k}  \nonumber \\ 
& \quad	\quad \quad \quad   \big|D_{[0:K-1]} = (d_0', \dots,d'_{k-1}, d_k , d'_{k+1}\dots  d'_{K-1})\Big),
\end{align*}
which means 
\begin{align*}  
& \left(X_{(d_0,\dots,d_{k-1},d_k,d_{k+1}, \dots, d_{K-1})},Z_{k},W_{d_k} \right)  \nonumber \\
\sim   & \left(X_{(d_0',\dots,d'_{k-1},d_k, d'_{k+1},\dots, d'_{K-1})},Z_{k},W_{d_k} \right).
\end{align*}
The proof is thus complete.

\section{Proof of Lemmas \ref{Lemcon3} and  \ref{Lemcon3b}} \label{pfLemcon3} 
\subsection{Preparation for Proof of Lemmas \ref{Lemcon3} and \ref{Lemcon3b}} 
Before we provide the proof of Lemmas \ref{Lemcon3} and \ref{Lemcon3b}, let's first prove the following lemma. 
\begin{Lem}\label{Lemcon2}
For all  $ a \in \{0,1\}$, $k\in[1:K]$, $ i \in [K-k:K-1], n\in[0:K-i-1]$, any coded caching scheme with $N=2$ satisfies the following inequalities,
% \begin{small} 
\begin{subequations}
	\begin{align} 	\label{Lemcon21}  % converse 1 使用
		\sum_{j=0}^{n} & H\left( Z_i,X_{(a\boldsymbol{1}_{K-j},\tilde{a} {\boldsymbol{1}}_{j})}\right)  
		\ge   n H(W_a,Z_i)   \nonumber \\
		& \quad \quad \quad  \quad  +    H\big( Z_i, \big ( X_{(a\boldsymbol{1}_{K-j},\tilde{a} {\boldsymbol{1}}_{j})} \big)_{j\in[0:n]} \big)   , \text{ and}  \\
		% -----------------------------------------------------
		\label{Lemcon22} % converse 2 使用
		\sum_{j=0}^{n} &  H\left( X_{(a\boldsymbol{1}_{K-i},\tilde{a} {\boldsymbol{1}}_{i})},Z_{j}\right) 
		\ge  nH \big(W_a,  X_{(a\boldsymbol{1}_{K-i},\tilde{a} {\boldsymbol{1}}_{i})}\big )   \nonumber \\
		& \quad \quad \quad \quad  + H\left(  X_{(a\boldsymbol{1}_{K-i},\tilde{a} {\boldsymbol{1}}_{i})},  Z_{[0:n]}  \right) .
	\end{align} 
\end{subequations}
% \end{small} 
\end{Lem}
Lemma \ref{Lemcon2}  provide ways to get rid of the summation in the sum of the joint entropies of a cache content and a delivery signal,  where \eqref{Lemcon21} deals with the case where the summation is over different delivery signals, and \eqref{Lemcon22} deals with the case where the summation is over different cache contents. 
% proof 符号改动
\begin{IEEEproof}
Firstly, we prove the inequality \eqref{Lemcon21}, which is done by induction. When $n=0$, both sides of the inequality are $H(Z_i,X_{a\boldsymbol{1}_{K}})$, and the inequality clearly holds.
Now suppose \eqref{Lemcon21} holds for  $n=m-1$,
we now prove that it is true for $n = m$ for $m\in[1:K-i-1] $.
%As \eqref{Lemcon21} holds for  $n=m-1$, 
Firstly, we have
\begin{align*}
& \textstyle \sum_{j=0}^{m} H\left( Z_i,X_{(a\boldsymbol{1}_{K-j},\tilde{a} {\boldsymbol{1}}_{j})} \right) \nonumber \\
%% --------------------------------------------------------
\overset{\text{(a)}}{\ge}   
& H(Z_i,X_{(a\boldsymbol{1}_{K-m},\tilde{a} {\boldsymbol{1}}_{m})} ) + (m-1) H(W_a,Z_i) \nonumber \\  
& + H\big( Z_i, 
\big( X_{(a\boldsymbol{1}_{K-j},\tilde{a} {\boldsymbol{1}}_{j})} \big)_{j\in[0:m-1]}\big) \\ 
%% --------------------------------------------------------
\overset{\text{(b)}}{\ge}   
& H(W_a, Z_i,X_{(a\boldsymbol{1}_{K-m},\tilde{a} {\boldsymbol{1}}_{m})} ) + (m-1) H(W_a,Z_i) \nonumber \\  
& + H\big(W_a, Z_i, 
\big(  X_{(a\boldsymbol{1}_{K-j},\tilde{a} {\boldsymbol{1}}_{j})} \big)_{j\in[0:m-1]}\big) \\ 
%% --------------------------------------------------------
\overset{\text{(c)}}{\ge}   
& m H(W_a,Z_i) + H\big( Z_i, 
\big( X_{(a\boldsymbol{1}_{K-j},\tilde{a} {\boldsymbol{1}}_{j})} \big)_{j\in[0:m]}\big) ,
\end{align*}
where  (a) follows from the supposition that the result holds true for $n=m-1$, 
(b) follows from the correctness constraint, i.e., \eqref{decoding}, 
and (c) follows from the submodularity of the entropy function stated in \eqref{submo}. The proof of \eqref{Lemcon21} is thus complete.

Next, we prove the inequality \eqref{Lemcon22}, again by induction. 
\begin{comment}
{\color{blue}For convenience of presentation, during the  proof of \eqref{Lemcon22}, let $X''_{(a,k,i)}$ denote $X_{(a\boldsymbol{1}_{k-i},\tilde{a} {\boldsymbol{1}}_{K-k+i})} $},  
\end{comment}
When $n=0$, both sides of the inequality are $ H( X_{(a\boldsymbol{1}_{K-i},\tilde{a} {\boldsymbol{1}}_{i})},Z_0)  $, and the inequality clearly holds.
Now  suppose \eqref{Lemcon22} holds for  $n=m-1$,
we now prove that it is true for $n = m$ for $m\in[1:K-i-1] $. Firstly, we have
%As  \eqref{Lemcon22} holds for  $n=m-1$, we have
%% 同样修改
\begin{align*}
& \textstyle \sum_{j=0}^{m} H\left( X_{(a\boldsymbol{1}_{K-i},\tilde{a} {\boldsymbol{1}}_{i})},Z_j\right) \nonumber \\
%% --------------------------------------------------------
\overset{\text{(a)}}{\ge}    &
H\left( X_{(a\boldsymbol{1}_{K-i},\tilde{a} {\boldsymbol{1}}_{i})} ,Z_m\right)  +  (m-1)H(W_a,X_{(a\boldsymbol{1}_{K-i},\tilde{a} {\boldsymbol{1}}_{i})})  
\nonumber \\
&    + H\left(  X_{(a\boldsymbol{1}_{K-i},\tilde{a} {\boldsymbol{1}}_{i})},  Z_{[0:m-1]}  \right)\\
%% --------------------------------------------------------
\overset{\text{(b)}}{\ge}    &
H\left(W_a, X_{(a\boldsymbol{1}_{K-i},\tilde{a} {\boldsymbol{1}}_{i})}, Z_m\right) 
+  (m-1)H(W_a,X_{(a\boldsymbol{1}_{K-i},\tilde{a} {\boldsymbol{1}}_{i})})  
\nonumber \\
&    + H\left(W_a, X_{(a\boldsymbol{1}_{K-i},\tilde{a} {\boldsymbol{1}}_{i})},  Z_{[0:m-1]}  \right)\\
%% --------------------------------------------------------
\overset{\text{(c)}}{\ge}  
& m H(W_a,X_{(a\boldsymbol{1}_{K-i},\tilde{a} {\boldsymbol{1}}_{i})})  
+ H\left(X_{(a\boldsymbol{1}_{K-i},\tilde{a} {\boldsymbol{1}}_{i})},  Z_{[0:m]}  \right),
\end{align*}
where  (a) follows from the supposition that the result holds true for $n=m-1$, 
(b) follows from the correctness constraint, i.e., \eqref{decoding}, 
and (c) follows from the submodularity of the entropy function stated in \eqref{submo}. The proof of \eqref{Lemcon22} is thus complete.
\end{IEEEproof}

Armed with the result of Lemma \ref{Lemcon2}, we are ready to prove Lemmas \ref{Lemcon3} and \ref{Lemcon3b}. 

\subsection{Proof of Lemma \ref{Lemcon3} } 
In this subsection, we present the proof of Lemma \ref{Lemcon3}. We have
\begin{subequations}
\begin{align} 
& \textstyle \sum_{i=K-k}^{K-1} \sum_{j=0}^{K-1-i} 
H\left( Z_i,X_{(a\boldsymbol{1}_{K-j},\tilde{a} \boldsymbol{1}_{j})}\right) \nonumber  \\
%% ---------------------------------------------------------------
% Z0应用lemma1  $n=k-2$ for $i=K-k$ 
\ge \label{pfLemcon31a} 
& H \left(Z_{K-k},X_{(a\boldsymbol{1}_{K-k+1},\tilde{a}\boldsymbol{1}_{k-1})}\right) 
+ (k-2) H(W_a,Z_{K-k})   \nonumber \\
& + H\big(Z_{K-k}, 		 (X_{(a\boldsymbol{1}_{K-j},\tilde{a}\boldsymbol{1}_{j})})_{j\in[0:k-2]}\big) \nonumber \\
% Z1-Z_{K-1} 用lemma 1 and $n=K-i-1$ for $i \in[K-k+1:K-1]$ 
& + \textstyle  \sum_{i=K-k+1}^{K-1}  \Big(  H\big( Z_i, ( X_{(a\boldsymbol{1}_{K-j},\tilde{a}\boldsymbol{1}_{j})})_{j\in[0:K-1-i]} \big)  \nonumber\\
&  \quad \quad \quad \quad \quad \quad \quad \quad + (K-1-i) H(W_a,Z_i)  \Big)  \\
%% -----------------------------------------------------
%   整理一下 decode
= \label{pfLemcon31b} 
& H \left(W_a, Z_{K-k},X_{(a\boldsymbol{1}_{K-k+1},\tilde{a}\boldsymbol{1}_{k-1})}\right)   \nonumber \\
& + (k-2) H(W_a,Z_{K-k})   \nonumber \\
& + \textstyle \sum_{i=K-k+1}^{K-2}(K-i-1) H(W_a,Z_i)  \nonumber \\
& + H\big(W_a, Z_{K-k}, 		   (X_{(a\boldsymbol{1}_{K-j},\tilde{a}\boldsymbol{1}_{j})})_{j\in[0:k-2]}\big)\nonumber \\
& + H\big(W_a, Z_{K-k+1}, 	(X_{(a\boldsymbol{1}_{K-j},\tilde{a}\boldsymbol{1}_{j})})_{j\in[0:k-2]}\big)
\nonumber \\
& + \textstyle  \sum_{i=K-k+2}^{K-1}  H\big( Z_i, ( X_{(a\boldsymbol{1}_{K-j},\tilde{a}\boldsymbol{1}_{j})})_{j\in[0:K-1-i]} \big)  \\
%% -----------------------------------------------------
\ge \label{pfLemcon31c} 
& H \left(W_a, Z_{K-k},X_{(a\boldsymbol{1}_{K-k+1},\tilde{a}\boldsymbol{1}_{k-1})}\right) \nonumber \\
& + (k-2) H(W_a,Z_{K-k})   \nonumber \\
& + \textstyle  \sum_{i=K-k+1}^{K-2}(K-i-1) H(W_a,Z_i)  \nonumber \\
%  (2) Z1Z2 合并 sub
& + H\big(W_a, Z_{K-k},  Z_{K-k+1}, 			(X_{(a\boldsymbol{1}_{K-j},\tilde{a}\boldsymbol{1}_{j})})_{j\in[0:k-2]}\big)\nonumber \\
& + H\big(W_a,	\big(X_{(a\boldsymbol{1}_{K-j},\tilde{a}\boldsymbol{1}_{j})} \big)_{j\in[0:k-2]}\big) 
\nonumber \\
& + \textstyle \sum_{i=K-k+2}^{K-1}  H\big( Z_i, ( X_{(a\boldsymbol{1}_{K-j},\tilde{a}\boldsymbol{1}_{j})})_{j\in[0:K-1-i]} \big)  \\
%% -----------------------------------------------------
\ge \label{pfLemcon31d} 
& \textstyle \sum_{i=K-k}^{K-2}(K-i-1) H(W_a,Z_i)  \nonumber \\
& + H\big(W_a, Z_{K-k},  Z_{K-k+1}, 			(X_{(a\boldsymbol{1}_{K-j},\tilde{a}\boldsymbol{1}_{j})})_{j\in[0:k-1]}\big)\nonumber \\
%  (2) 用递推
& + (k-2)H(W_{a}, W_{\tilde{a}}) +   H\left(W_a, X_{a\boldsymbol{1}_{K}}\right) \\
%% ----------------------------------------------------- 
\ge \label{pfLemcon31e} 
% \textstyle
& \textstyle  \sum_{i=K-k}^{K-2}(K-i-1) H(W_a,Z_i)  \nonumber \\  
&   +  (k-1)H(W_{a}, W_{\tilde{a}})  +  H\left(W_a, X_{a\boldsymbol{1}_{K}}\right) \\  
= \label{pfLemcon31f} 
& \textstyle \sum_{i=K-k}^{K-2}(K-i-1) H(W_a,Z_i)   \nonumber \\  
&  +  2(k-1)F + H\left(W_a, X_{a\boldsymbol{1}_{K}}\right) ,
\end{align}
\end{subequations}
where \eqref{pfLemcon31a}  follows from the result of Lemma \ref{Lemcon2}, more specifically, we set $n=k-2$ for $i=K-k$ and $n=K-i-1$ for $i \in[K-k+1:K-1]$ in  \eqref{Lemcon21},
\eqref{pfLemcon31b} and  \eqref{pfLemcon31e} follow from the correctness constraint, i.e., \eqref{decoding},
\eqref{pfLemcon31c}  and the first two terms in  \eqref{pfLemcon31d}  follow from the submodularity of the entropy function stated in \eqref{submo},
and \eqref{pfLemcon31f} follows from the assumption that the files are independent, uniformly distributed   and  $F$ bits each.
The last two terms in  \eqref{pfLemcon31d}   follow from the recursive proof below:
\begin{align*}
& H\big(W_a,	 (X_{(a\boldsymbol{1}_{K-j},\tilde{a}\boldsymbol{1}_{j})}  )_{j\in[0:k-2]}\big) 
\nonumber \\
& + \textstyle  \sum_{i=K-k+2}^{K-1}  H\big( Z_i, ( X_{(a\boldsymbol{1}_{K-j},\tilde{a}\boldsymbol{1}_{j})})_{j\in[0:K-1-i]} \big) \\
%% ----------------------------------------------------- 
\overset{\text{(a)}}{=} &   % 递推第一步
H\big(W_a,	 (X_{(a\boldsymbol{1}_{K-j},\tilde{a}\boldsymbol{1}_{j})}  )_{j\in[0:k-2]}\big) \nonumber \\
& + H\big(W_a,Z_{K-k+2}, (X_{(a\boldsymbol{1}_{K-j},\tilde{a}\boldsymbol{1}_{j})})_{j\in[0:k-3]} \big) \nonumber \\
& + \textstyle \sum_{i=K-k+3}^{K-1}  H\big( Z_i, ( X_{(a\boldsymbol{1}_{K-j},\tilde{a}\boldsymbol{1}_{j})})_{j\in[0:K-1-i]} \big) \\
%% ----------------------------------------------------- 
%  submodularity of entropy functional 
\overset{\text{(b)}}{\ge }  % 递推第二步
& H\big(W_a,Z_{K-k+2},  (X_{(a\boldsymbol{1}_{K-j},\tilde{a}\boldsymbol{1}_{j})}  )_{j\in[0:k-2]}\big) \nonumber \\
& + H\big(W_a, (X_{(a\boldsymbol{1}_{K-j},\tilde{a}\boldsymbol{1}_{j})})_{j\in[0:k-3]} \big) \nonumber \\
& +  \textstyle \sum_{i=K-k+3}^{K-1}  H\big( Z_i, ( X_{(a\boldsymbol{1}_{K-j},\tilde{a}\boldsymbol{1}_{j})})_{j\in[0:K-1-i]} \big) \\
%% ----------------------------------------------------- 
%  submodularity of entropy functional 
\overset{\text{(c)}}{\ge }  % 递推第三步
& H(W_{a}, W_{\tilde{a}})
+ H\big(W_a, (X_{(a\boldsymbol{1}_{K-j},\tilde{a}\boldsymbol{1}_{j})})_{j\in[0:k-3]} \big) \nonumber \\
& + H\big(W_a, Z_{K-k+3}, ( X_{(a\boldsymbol{1}_{K-j},\tilde{a}\boldsymbol{1}_{j})})_{j\in[0:k-4]} \big) \nonumber \\ 
& + \textstyle \sum_{i=K-k+4}^{K-1}  H\big( Z_i, ( X_{(a\boldsymbol{1}_{K-j},\tilde{a}\boldsymbol{1}_{j})})_{j\in[0:K-1-i]} \big) \\
%% ----------------------------------------------------- 
\ge & \dots  \overset{\text{(d)}}{\ge} % 递推第四步
(k-3)H(W_{a}, W_{\tilde{a}}) + H\big(W_a, (X_{(a\boldsymbol{1}_{K-j},\tilde{a}\boldsymbol{1}_{j})})_{j\in[0:1]} \big) \\
& +   H\left(W_a,  Z_{K-1}, X_{a\boldsymbol{1}_{K}}  \right)  \\
%% ----------------------------------------------------- 
\overset{\text{(e)}}{\ge }  % 递推第五步
&  (k-3)H(W_{a}, W_{\tilde{a}}) +  H\left(W_a, X_{a\boldsymbol{1}_{K}}  \right)   \\
&  + H\big(W_a, Z_{K-1}, (X_{(a\boldsymbol{1}_{K-j},\tilde{a}\boldsymbol{1}_{j})})_{j\in[0:1]} \big) \\
%% ----------------------------------------------------- 
\overset{\text{(f)}}{\ge }  % 递推第六步
&  (k-2)H(W_{a}, W_{\tilde{a}})
+  H\left(W_a, X_{a\boldsymbol{1}_{K}}  \right),
\end{align*}
% 递推第一步原因 
where  $(a)$, $(c)$ and $(f)$ follow from the correctness constraint, i.e., \eqref{decoding}, 
$(b)$ and $(e)$ follow from the submodularity of the entropy function stated in \eqref{submo}, 
and  $(d)$ follows from  repeating steps $(b)$ and $(c)$.

The proof of Lemma \ref{Lemcon3} is thus complete. 

\subsection{Proof of Lemma \ref{Lemcon3b} } 
In this subsection, we present the proof of Lemma \ref{Lemcon3b}. We have
% dual 
\begin{subequations}
\begin{align} 
& \textstyle \sum_{i=K-k}^{K-1} \sum_{j=0}^{K-1-i}  H\left( 	X_{(a\boldsymbol{1}_{K-i},\tilde{a} \boldsymbol{1}_{i})},Z_j\right) \nonumber \\
%% -----------------------------------------------
% Z0 lemma 1 % $n=k-2$ for $i=K-k$ and 
\ge \label{pfLemcon32a} 
&  H(X_{(a\boldsymbol{1}_{k},\tilde{a} \boldsymbol{1}_{K-k})},Z_{k-1})  + (k-2) H(W_a,X_{(a\boldsymbol{1}_{k},\tilde{a} \boldsymbol{1}_{K-k})})  \nonumber \\  
& +  H\left(X_{(a\boldsymbol{1}_{k},\tilde{a} \boldsymbol{1}_{K-k})}, Z_{[0:k-2]}\right)   \nonumber \\
% Z1-Z_{K-1} 用lemma 1 $n=K-i-1$ for $i \in[K-k+1:K-1]$
& +  \textstyle \sum_{i=K-k+1}^{K-1}  \Big( 
H\left(X_{(a\boldsymbol{1}_{K-i},\tilde{a}\boldsymbol{1}_{i})}, Z_{[0:K-1-i]}
\right)  \nonumber \\
& \quad \quad \quad \quad \quad + (K-1-i) H(W_a,X_{(a\boldsymbol{1}_{K-i},\tilde{a}\boldsymbol{1}_{i})}) \Big)  \\
%%  ------------------------------------- 
= \label{pfLemcon32b} % 解码
&  H(W_a, X_{(a\boldsymbol{1}_{k},\tilde{a} \boldsymbol{1}_{K-k})},Z_{k-1}) \nonumber \\ 
&  + (k-2) H(W_a,X_{(a\boldsymbol{1}_{k},\tilde{a} \boldsymbol{1}_{K-k})})  \nonumber \\  
& + \textstyle \sum_{i=K-k+1}^{K-2} (K-1-i) H(W_a,X_{(a\boldsymbol{1}_{K-i},\tilde{a}\boldsymbol{1}_{i})}) \nonumber \\ 
% sub 递推
& +  H\left(W_a, X_{(a\boldsymbol{1}_{k},\tilde{a} \boldsymbol{1}_{K-k})}, Z_{[0:k-2]}\right)   \nonumber \\
& +  H\left(W_a, X_{(a\boldsymbol{1}_{k-1},\tilde{a} \boldsymbol{1}_{K-k+1})}, Z_{[0:k-2]}\right)   \nonumber \\
& + \textstyle \sum_{i=K-k+2}^{K-1}  
H\left(X_{(a\boldsymbol{1}_{K-i},\tilde{a}\boldsymbol{1}_{i})}, Z_{[0:K-1-i]}
\right)  \\
%%  ------------------------------------- 
\ge \label{pfLemcon32c} % sub
&  H(W_a, X_{(a\boldsymbol{1}_{k},\tilde{a} \boldsymbol{1}_{K-k})},Z_{k-1}) \nonumber \\ 
&  + (k-2) H(W_a,X_{(a\boldsymbol{1}_{k},\tilde{a} \boldsymbol{1}_{K-k})})  \nonumber \\  
& + \textstyle \sum_{i=K-k+1}^{K-2} (K-1-i) H(W_a,X_{(a\boldsymbol{1}_{K-i},\tilde{a}\boldsymbol{1}_{i})}) \nonumber \\ 
% sub 递推
& +  H\left(W_a, X_{(a\boldsymbol{1}_{k},\tilde{a} \boldsymbol{1}_{K-k})},X_{(a\boldsymbol{1}_{k-1},\tilde{a} \boldsymbol{1}_{K-k+1})}, Z_{[0:k-2]}\right)   \nonumber \\
& +  H\left(W_a,  Z_{[0:k-2]}\right)   \nonumber \\
& +  \textstyle \sum_{i=K-k+2}^{K-1}  
H\left(X_{(a\boldsymbol{1}_{K-i},\tilde{a}\boldsymbol{1}_{i})}, Z_{[0:K-1-i]} \right)  \\
%%  ------------------------------------- 
\ge \label{pfLemcon32d} 
& \textstyle \sum_{i=K-k}^{K-2} (K-1-i) H(W_a,X_{(a\boldsymbol{1}_{K-i},\tilde{a}\boldsymbol{1}_{i})}) \nonumber \\ 
& +  H\left(W_a, X_{(a\boldsymbol{1}_{k},\tilde{a} \boldsymbol{1}_{K-k})},X_{(a\boldsymbol{1}_{k-1},\tilde{a} \boldsymbol{1}_{K-k+1})}, Z_{[0:k-1]}\right)   \nonumber \\
% 应用递推得到的结果
& + (k-2) H(W_a,W_{\tilde{a}}) + H\left(W_a, Z_{0}\right)\\
%%  -------------------------------------------------------
\ge \label{pfLemcon32e} 
& \textstyle \sum_{i=K-k}^{K-2} (K-1-i) H(W_a,X_{(a\boldsymbol{1}_{K-i},\tilde{a}\boldsymbol{1}_{i})}) \nonumber \\ 
& + (k-1) H(W_a,W_{\tilde{a}}) + H\left(W_a, Z_{0}\right)\\
%%  -------------------------------------------------------
=   \label{pfLemcon32f} 
& \textstyle \sum_{i=K-k}^{K-2} (K-1-i) H(W_a,X_{(a\boldsymbol{1}_{K-i},\tilde{a}\boldsymbol{1}_{i})}) \nonumber \\ 
& + 2(k-1)F + H\left(W_a, Z_{0}\right),
\end{align}
\end{subequations}
where \eqref{pfLemcon32a} follows from the result of Lemma \ref{Lemcon2}, more specifically, we set $n=k-2$ for $i=K-k$ and $n=K-i-1$ for $i \in[K-k+1:K-1]$ in  \eqref{Lemcon22}, 
\eqref{pfLemcon32b} and  \eqref{pfLemcon32e} follow from the correctness constraint, i.e., \eqref{decoding}, 
\eqref{pfLemcon32c}  and the first two terms in  \eqref{pfLemcon32d}  follow from the submodularity of the entropy function stated in \eqref{submo},
and \eqref{pfLemcon32f} follows from the assumption that the files are independent, uniformly distributed   and  $F$ bits each.
The last two terms in \eqref{pfLemcon32d} follow from the recursive proof below:
\begin{align*}
& H\left(W_a,  Z_{[0:k-2]}\right)   \nonumber \\
& +  \textstyle \sum_{i=K-k+2}^{K-1}  
H\left(X_{(a\boldsymbol{1}_{K-i},\tilde{a}\boldsymbol{1}_{i})}, Z_{[0:K-1-i]} \right)  \\
%% ----------------------------------------------------- 
\overset{\text{(a)}}{=}    % 递推第一步
& H\left(W_a,  Z_{[0:k-2]}\right)  + H\left(W_a, X_{(a\boldsymbol{1}_{k-2},\tilde{a}\boldsymbol{1}_{K-k+2})}, Z_{[0:k-3]} \right) \nonumber \\
& +  \textstyle \sum_{i=K-k+3}^{K-1}  
H\left(X_{(a\boldsymbol{1}_{K-i},\tilde{a}\boldsymbol{1}_{i})}, Z_{[0:K-1-i]} \right)  \\
%% ----------------------------------------------------- 
\overset{\text{(b)}}{\ge}    %  递推第二步
& H\left(W_a, X_{(a\boldsymbol{1}_{k-2},\tilde{a}\boldsymbol{1}_{K-k+2})}, Z_{[0:k-2]} \right) + H\left(W_a,  Z_{[0:k-3]}\right)  \nonumber \\
& +  \textstyle \sum_{i=K-k+3}^{K-1}  
H\left(X_{(a\boldsymbol{1}_{K-i},\tilde{a}\boldsymbol{1}_{i})}, Z_{[0:K-1-i]} \right)  \\
%% ----------------------------------------------------- 
\overset{\text{(c)}}{\ge}  % 递推第三步
&  H(W_{a}, W_{\tilde{a}})  + H\left(W_a,  Z_{[0:k-3]}\right)  \nonumber \\
& +  H\left(W_a, X_{(a\boldsymbol{1}_{k-3},\tilde{a} \boldsymbol{1}_{K-k+3})}, Z_{[0:k-4]}\right) \\
& +  \textstyle \sum_{i=K-k+4}^{K-1}  
H\left(X_{(a\boldsymbol{1}_{K-i},\tilde{a}\boldsymbol{1}_{i})}, Z_{[0:K-1-i]} \right)  \\
%% ----------------------------------------------------- 
\ge & \dots  \overset{\text{(d)}}{\ge}     % 递推第四步
(k-3)H(W_{a}, W_{\tilde{a}})  + H\left(W_a, Z_{[0:1]} \right)  
\\
& + H\left(W_a, X_{(a,\tilde{a}\boldsymbol{1}_{K-1})}, Z_0\right)  \\
%% ----------------------------------------------------- 
\overset{\text{(e)}}{\ge } & % 递推第五步
(k-3)H(W_{a}, W_{\tilde{a}})  + H\left(W_a, Z_{[0:1]}, X_{(a,\tilde{a}\boldsymbol{1}_{K-1})}\right)   \\
& + H\left(W_a, Z_0\right)  \\
%% ----------------------------------------------------- 
\overset{\text{(f)}}{\ge }  % 递推第六步
&  (k-2)H(W_{a}, W_{\tilde{a}})
+ H\left(W_a, Z_0\right), 
\end{align*}
% 递推第一步原因 
where  $(a)$, $(c)$ and $(f)$ follow from the correctness constraint, i.e., \eqref{decoding}, 
$(b)$ and $(e)$ follow from the submodularity of the entropy function stated in \eqref{submo}, 
and $(d)$ follows from  repeating steps $(b)$ and $(c)$.

The proof of Lemma \ref{Lemcon3b} is thus complete.

\vspace{3mm}

\section{Proof of Corollaries \ref{cor1} and \ref{cor2}} \label{pfCoro}  
\subsection {Proof of Corollary \ref{cor1} } \label{pfCoro1}
From Theorem \ref{ach2}, let $N =2$ and  $r=K+1,K,K-1$,  the $(M,R)$ given by $(0,2), (\frac{1}{K+1},\frac{2K}{K+1})$ and $ (\frac{2}{K},\frac{2(K-1)}{K+1} )$ 
are achievable.
From \cite[Theorem 2]{Chinmay2022}, let $N =2$ and  $r=K+1,K,K-1$,  the $(M,R)$ given by $(2,0), (\frac{2K}{K+1},\frac{1}{K+1} )$ and $ (\frac{2(K-1)}{K+1},\frac{2}{K} )$ are achievable.
The lower convex envelope of the above memory-rate pairs can be achieved by memory-sharing.

Next, from Theorem \ref{con2}, let $k=K$, we get 
\begin{align*}
(K+1)(K+2) M + 2K(K+1)R &\ge 2K(K+3), \text{ and }   \\
2K(K+1)M + (K+1)(K+2) R &\ge 2K(K+3)  .
\end{align*}
% we set  $N=2,k=K$ in Theorem \ref{con2} and $N=2, s=1,2$ in \cite[Theorem 2]{MaddahAli2014}, thus for the $(2,K)$ demand private coded caching problem, any $(M, R)$ pair {\color{red}must satisfy}
From \cite[Theorem 2]{MaddahAli2014}, which is a converse result for the traditional coded caching problem not considering privacy, let $N=2, s=1,2$, we get $2M + R \ge 2$ and $M+2R \ge 2$. 
Thus, for the $(2,K)$ demand private coded caching problem, any $(M, R)$ pair must satisfy
\begin{align*} 
R^{*p}_{N,K}(M) \ge & \max \Big\{2-2M, \frac{2K(K+3)}{(K+1)(K+2)}-\frac{2K}{K+2}M ,\nonumber  \\ 
& \quad \quad \quad \quad  \quad \frac{K+3}{K+1}-\frac{K+2}{2K}M ,1- \frac{1}{2}M
\Big\}.
\end{align*}
The above converse and achievability results meet when $M\in [0,\frac{2}{K}] \cup [\frac{2(K-1)}{K+1},2]$.

The proof of Corollary \ref{cor1} is thus complete.

\subsection {Proof of Corollary \ref{cor2} } \label{pfCoro2}
From Theorem \ref{con2}, let $k=2,3$, we get $6M + 5R \ge 9 $, $ 5M + 6R \ge 9 $ and $3M + 3R \ge 5$. 
From \cite[Theorem 2]{MaddahAli2014}, which is a converse result for the traditional coded caching problem not considering privacy, let $N=2, s=1,2$, we get $2M + R \ge 2$ and $M+2R \ge 2$. 
Combine the above converse results, for the $(2,3)$ demand private coded caching problem, any $(M, R)$ pair must satisfy
\begin{align*} 
& R_{N,K}^{*p}(M)   \\
& \quad \ge \max \Big \{ 2-2M, \frac{9-6M}{5}, \frac{5-3M}{3},  \frac{9-5M}{6} ,  \frac{2-M}{2} \Big \}.
\end{align*}
% The six corner points, i.e., $(0,2), (\frac{1}{4},\frac{3}{2}),(\frac{2}{3},1),(1,\frac{2}{3}) ,(\frac{3}{2},\frac{1}{4})$ and $(2,0)$ can be obtained by setting $N =2, K=3$ and  $r=4,3,2$ in Theorem \ref{ach2} and \cite[Theorem 2]{Chinmay2022}.
The six corner points, i.e., $(0,2), (\frac{1}{4},\frac{3}{2}),(\frac{2}{3},1),(1,\frac{2}{3}) ,(\frac{3}{2},\frac{1}{4})$ and $(2,0)$ can be obtained from the following schemes.  
From Theorem \ref{ach2}, let $N =2, K=3$ and  $r=4,3,2$,  the $(M,R)$ given by $(0,2), (\frac{1}{4},\frac{3}{2})$ and $(\frac{2}{3},1)$ are achievable.
From \cite[Theorem 2]{Chinmay2022}, let $N =2, K=3$ and  $r=4,3,2$,  the $(M,R)$ given by $(2,0)$, $(\frac{3}{2},\frac{1}{4})$ and $(1,\frac{2}{3})$ are achievable.
% by setting $N =2, K=3$ and  $r=4,3,2$ in  and \cite[Theorem 2]{Chinmay2022}.

The proof of Corollary \ref{cor2} is thus complete.
%\bibliographystyle{IEEEtran}
%\bibliography{IEEEexample}

		% %%%%%%%%%%%%%%%%%%%%%%%%%%%%%%%%%%%%%%%%%
		% \begin{figure}[t!]
		%\centering
		%\includegraphics[width=3.5in]{../Figures/P10a09b09_final}
		%\caption{The achievable sum rate vs. capacity of the backhaul link for $P=10$, $a=b=0.9$ and $C_1=C_2=C$.} \label{final_plot}
		%\end{figure}
		%%%%%%%%%%%%%%%%%%%%%%%%%%%%%%%%%%%%%%%%%
		% \clearpage
		% \balance
		%\bibliographystyle{unsrt}
		%\bibliographystyle{IEEEtran}
		%\bibliography{definitions,bibliofile}
		%\bibliographystyle{ieeetr}
		%\bibliography{ref}
		
		%\begin{thebibliography}{1}
		\bibliographystyle{IEEEtran}
		\bibliography{ref}
		%\end{thebibliography}

		% \balance

	\end{document}